\documentclass[aps,twocolumn,pre,noshowpacs,superscriptaddress,showkeys,floatfix]{revtex4}

\usepackage{amsthm,amsmath}
\RequirePackage{natbib}
\usepackage[utf8]{inputenc} 

\usepackage{graphicx}
\usepackage{array}

\newcommand{\fref}[1]{Figure~#1}
\newcommand{\freftwo}[2]{Figures~#1 and #2}
\newcommand{\frefs}[1]{Figures~#1}
\newcommand{\sfref}[1]{Additional file~#1: Figure~S#1}
\newcommand{\sfrefs}[1]{Additional file~#1: Figures~S#1}
\newcommand{\supmethods}[0]{Additional file~15: Supplementary Methods}
\newcommand{\stref}[1]{{Additional file~\number\numexpr#1+12\relax: Table~S#1}}
\newcommand{\movref}[1]{{Additional file~\number\numexpr#1+15\relax: Video~S#1}}
\newcommand{\matmeth}[0]{Methods}
\newcommand{\sfreftwo}[2]{Figures~S#1 and S#2}

\renewcommand{\figurename}{\textbf{Figure} }
\renewcommand{\thefigure}{\textbf{\arabic{figure}}}

\usepackage{titlesec}
\titleformat*{\section}{\large\bfseries\centering}

\newcommand{\citelangowski}{1,55}  
\newcommand{\citeHughes}{14}
\newcommand{\citeMarques}{56}
\newcommand{\citeKassouf}{50}
\newcommand{\citeTallack}{57}
\newcommand{\citeENCODE}{58}

\makeatletter
\newcommand*{\balancecolsandclearpage}{%
  \close@column@grid
  \clearpage
  \twocolumngrid
}
\makeatother

\begin{document}

\title{Predicting the three-dimensional folding of \textit{cis}-regulatory regions in mammalian genomes using bioinformatic data and polymer models}

\author{Chris A Brackley}
\affiliation{SUPA, School of Physics and Astronomy, University of Edinburgh}

\author{Jill M Brown} 
\affiliation{MRC Molecular Haematology Unit, Weatherall Institute Of Molecular Medicine, Oxford University}

\author{Dominic Waithe}
\affiliation{Wolfson Imaging Centre Oxford, Weatherall Institute Of Molecular Medicine, Oxford University}

\author{Christian Babbs}
\author{James Davies}
\author{Jim R Hughes}
\author{Veronica J Buckle}
\affiliation{MRC Molecular Haematology Unit, Weatherall Institute Of Molecular Medicine, Oxford University}

\author{Davide Marenduzzo}
\affiliation{SUPA, School of Physics and Astronomy, University of Edinburgh}

\begin{abstract} 
The three-dimensional organisation of chromosomes can be probed using methods such as Capture-C. However it is unclear how such population level data relates to the organisation within a single cell, and the mechanisms leading to the observed interactions are still largely obscure. We present a polymer modelling scheme based on the assumption that chromosome architecture is maintained by protein bridges which form chromatin loops. To test the model we perform FISH experiments and also compare with Capture-C data. Starting merely from the locations of protein binding sites, our model accurately predicts the experimentally observed chromatin interactions, revealing a population of 3D conformations.
\end{abstract}


\maketitle

\section*{Background}

The three-dimensional spatial organisation of mammalian chromosomes {\it in vivo} is a topic of fundamental importance in cell biology~\cite{3c,hic,hicdixon,hic14,Bickmore2013}.  Understanding how chromatin conformation becomes modified on a local scale in order to up-regulate transcription from genes during differentiation or development is critical not only to decipher a fundamental biological process, but also to delineate the role this process may play in human disease and potential therapies. The higher scale organisation of chromatin in the nucleus also has important roles to play in this regard~\cite{Bickmore2013,factories,misteli,aging,misteliaging} as the spatial structure of chromosomes is tightly linked to transcription. For instance, active genes can cluster at nuclear speckles~\cite{Shopland2003,Brown2008}; conversely peripheral lamina-associated domains (LADs) comprise of regions of the DNA that are not generically transcriptionally active~\cite{Guelen2008,Kind2013}. The three dimensional structure of the genome is, therefore, intimately related to its function.

Thanks to the development of high-throughput experimental techniques based on chromosome conformation capture (3C)~\cite{3c}, such as Hi-C and Capture-C~\cite{hic,hicdixon,Hughes,hic14,tetheredhic}, it is now possible to probe experimentally which regions of the genome of a given cell type are spatially proximate {\it in vivo}. 
A major result obtained with these methods has been the discovery that chromosomes are organised in a series of topologically-associated domains (TADs)~\cite{hic,hicdixon,hic14}, which are separated by boundaries, but whose biological nature remains elusive. While the TAD boundaries are thought to be largely conserved across cell types, the arrangement of the chromatin within a TAD is not~\cite{Nora2012}. This internal organisation depends on the activity of the genes within a domain, and is likely related to the action of \textit{cis}-regulatory elements (DNA regions where the binding of a transcription factor (TF) can regulate the expression of a gene which is tens or hundreds of kilo-base-pairs (kbp) away)~\cite{Lettice2002,Seawright2006}.

The pattern of interactions revealed by most 3C based experiments is an average over a large population of cells, yet it has become clear that there is a remarkable variability in both chromosomal conformation and chromatin interactions between different cells~\cite{frasersinglecells,tiana}. Thus it is an important challenge to understand how the chromosome conformation in single cells leads to the observed population average, and to decipher the mechanism underlying such arrangements. To address this issue, here we present an {\it in silico} investigation of the local folding and resulting interaction maps of important active gene loci in mouse erythroblasts. We concentrate on the well studied $\alpha$ and $\beta$ globin loci which have long been model systems for understanding \textit{cis}-regulatory interactions~\cite{Trimborn1999,Tolhuis2002,Anguita2002,Palstra2003,Splinter2006,Higgs2008,Vernimmen2009,Hou2010,Bau2011,Junier2012,Hughes}. These loci are known to have tissue-specific organisation, and expression of the different genes within the loci varies through development and erythropoiesis. As a comparison, we also study embryonic stem cells where these genes are not active. Our main result is that our model predicts patterns of contacts which are close to that found by high-resolution Capture-C experiments, reproduces the changes in such patterns following differentiation, and explains existing observations on the biology of the globin loci in mouse. Our predictions also compare favourably with new fluorescence in situ hybridisation (FISH) experiments that give spatial separation measurements between specific genomic locations in individual cells. This level of agreement is especially remarkable because it essentially involves no fitting. 

Our model builds on the minimal assumption that the spatial organisation of eukaryotic chromosomes is maintained largely through the action of proteins or protein complexes which can form bridges by simultaneously binding to more than one site in the genome, and forming loops from the intervening chromatin~\cite{nicodemi1,nicodemi2,bridging1,bridging2,hic14,Dowen2014,Lesne2014}. We treat the chromatin fibre as a simple bead-and-spring polymer (\sfref{1}), and coarse-grain the bridge forming protein complexes into single units. We then ``paint'' the polymer according to bioinformatic data characterising protein binding and chromatin state in the relevant cell type, and use molecular dynamics to simulate the motion of the region of interest (see \sfref{1} for a schematic diagram, and \supmethods{} for the full details of the model). The chromatin fibre and proteins diffuse as though subject to the thermal fluctuations of the nucleoplasm; the protein complexes can bind and dissociate from the chromatin and form bridges, and the fibre adopts conformations which are consistent with the entropic and energetic constraints of the system. By repeatedly running the simulation with different random thermal motions, we can generate a population of equilibrium conformations representing a population of cells. Some examples of other studies where polymer models have been applied to study chromatin are~\cite{tiana,nicodemi1,nicodemi2,bridging1,bridging2,Tark-Dame2014,Pombo2014,Doyle2014,Mirny2015}.

To keep our model as simple as possible we use the locations of DNase1 hypersensitive sites (DHSs) as a proxy for binding sites of a generic type of protein bridge, which we imagine is made up from complexes of transcription factors and other DNA-binding proteins. The choice of DHSs as binding sites is justified due to their well documented tendency to correlate with open chromatin, or euchromatin, and with peaks in ChIP-seq data for many transcription factors~\cite{Thurman2012}, such as GATA1, Nfe2 Scl/Tal1 and Klf1, all of which are known to be important for globin regulation (see \sfref{2}). The interactions between the many transcription factors and co-factors which might form the bridging complexes involved in \textit{cis}-regulatory binding are not well characterized, and the DHS approximation avoids the need to make any assumptions. One factor that most certainly has a chromatin architectural role is the CCCTC-binding factor (CTCF)~\cite{Holwerda2013,hic14,Dowen2014,Mirny2015,deWit2015,Sanborn2015}. This protein is thought to form dimers which drive looping between some of its specific binding sites scattered along the chromosomes of eukaryotic organisms. In particular, convergent CTCF binding sites have been proposed to delimit the extent of chromatin domains, which might be extruded through a looping complex, possibly comprising cohesin~\cite{Mirny2015,Sanborn2015,Imakaev2015}. CTCF is therefore a bridge with an architectural role, and has indeed been dubbed a ``global genome organizer''~\cite{Holwerda2013,hic14,Dowen2014}. Interestingly, chromatin has been found to compact on depletion of RAD21 and CTCF~\cite{Tark-Dame2014}. To reflect its perceived importance, we treat CTCF proteins as separate bridges in the simulations; in this case the binding sites are placed at peaks in the ChIP-seq data for CTCF binding (see \sfref{2}). Our model therefore includes two species of putative protein bridges, which we denote CTCF and DHS binding proteins (or bridges) respectively. Furthermore, we consider the hypothesis that some histone modifications (e.g. H3K4 monomethylation at enhancers or trimethylation at active promoters) act to recruit bridging proteins~\cite{Calo2013}. We include this in the model by introducing a weaker, non-specific interaction between the bridges and H3K4me1 modified regions (which are not already labelled as CTCF or DHS bridges); since the hypersensitive sites at regulatory elements are often surrounded by H3K4me1 modified regions, these act as a funnel which effectively directs proteins to their high affinity binding sites~\cite{Brackley2012}.

\section*{Results}

\subsection*{Chromatin folding in the mouse $\boldsymbol{\alpha}$ globin locus}

\begin{figure*}[th!]
\includegraphics{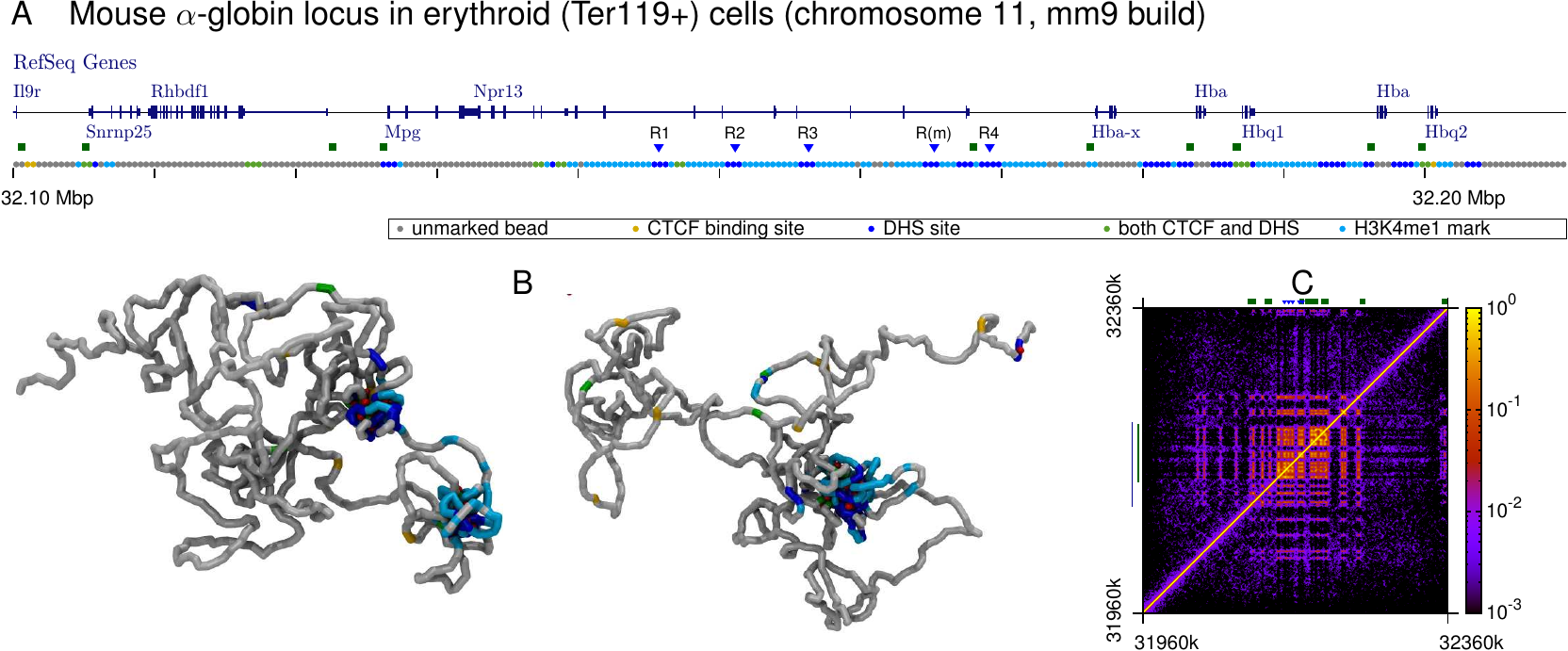}  
\caption{\textbf{Simulating the $\boldsymbol{\alpha}$ globin locus.} (A) Browser view showing genes in the vicinity of the $\alpha$ globin locus, alongside a schematic indicating the coarse-graining used in the simulations. A 110~kbp section of the 400~kbp chromatin fragment which was simulated is shown. As described in the text, simulation chromatin beads were designated as CTCF binding sites, DHS binding, H3K4me1 modified sites, and combinations of these. The positions of the set of five regulatory elements are indicated with blue triangles, and promoters with green squares. (B) Example simulated configurations of the locus. CTCF proteins (green) and DHS binding proteins (red) are shown; the chromosome fragment is coloured as in A. See also \movref{1} for a 3-D view of the configurations. Parameters for the polymer model and the bridge--chromatin affinity are given in full in \supmethods{}. (C) Contact map showing the frequency of contacts between each chromatin bead in 1000 simulated configurations. Note that the colour bar shows a logarithmic scale. The blue line to the left indicates the region which is shown in A. The green line to the left indicates the region which is used for the clustering analysis (\fref{2} and text). }
\end{figure*}

First, we use our model to predict the folding of a 400~kbp region around the mouse $\alpha$ globin locus (chr11:31960000-32360000, mm9 build; each polymer bead represents 400~bp, or two nucleosomes, see \fref{1A} and \matmeth{}). This well studied cluster contains five globin related genes: the $\zeta$ globin gene (\textit{Hba-x}, expressed in embryonic erythroid cells, but silent in adult cells), two copies of the $\alpha$ globin gene (\textit{Hba}, expressed in foetal and adult erythroblasts) and two $\theta$ globin genes (\textit{Hbq1} and \textit{Hbq2}, only weakly expressed in adult tissue). Expression of the genes in the cluster is controlled by several regulatory elements: the multi-species conserved elements R1-4 and the mouse specific R(m). Some of these are contained within the introns of \textit{Nprl3}, one of several widely expressed genes which surround the locus; the R2 element (known as HS-26 in mouse and equivalent to HS-40 in human) is thought to be particularly important for globin regulation~\cite{Trimborn1999,Anguita2002,Vernimmen2009}. \fref{1A} shows the binding sites for CTCF and DHS across the region considered (informed by ChIP-seq and DNase-seq data for adult erythroid cells -- see \sfref{2}); the positions of the H3K4me1 methylation marks are also indicated (from ChIP-seq data for the same cell type, see \sfref{2}). In our simulations, proteins bind strongly to the CTCF or DHS labelled beads, and also weakly to the H3K4me1 marks. Some typical snapshots from our simulations are shown in \fref{1B} and \movref{1} (CTCF and DHS binding proteins are shown as red and green spheres respectively), while the average contact map is shown in \fref{1C}.

\begin{figure}[th!]
\includegraphics[width=0.475\textwidth]{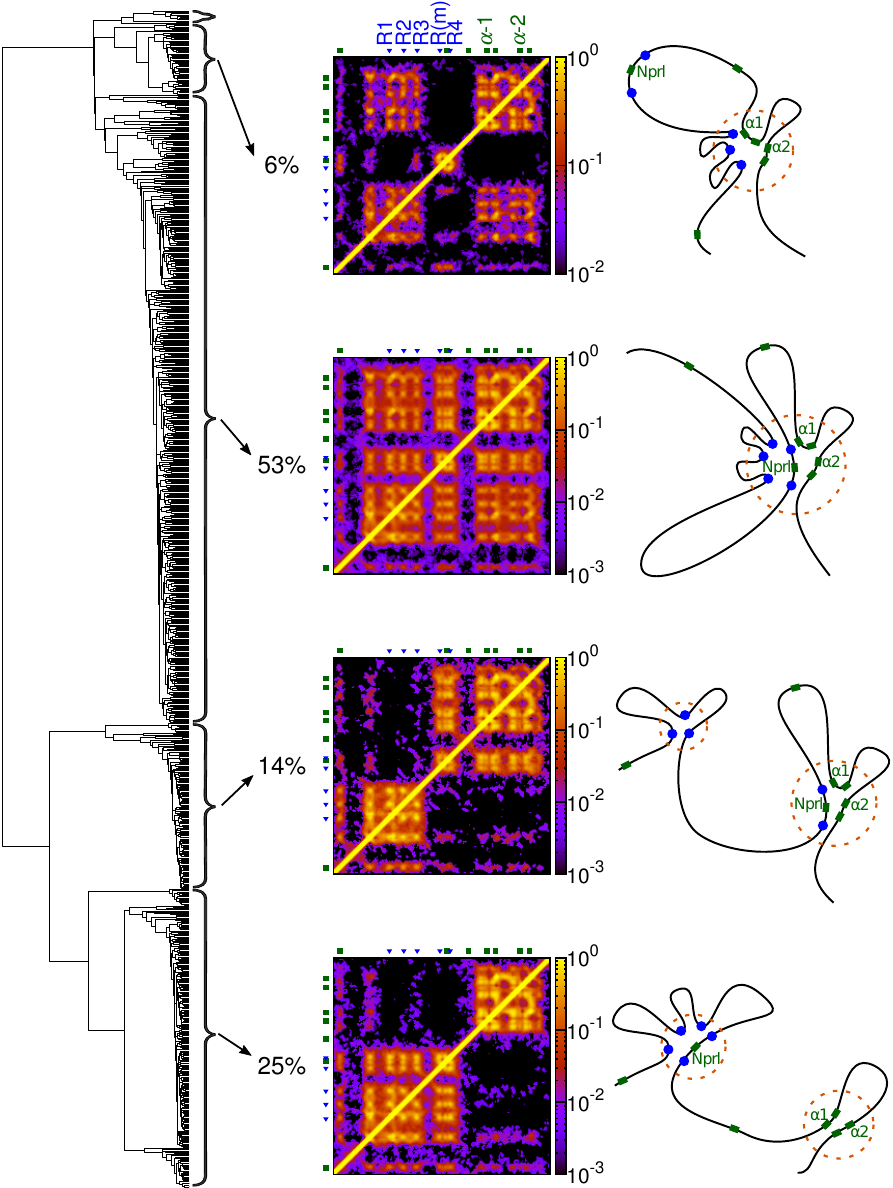}  
\caption{\textbf{Conformations of the $\boldsymbol{\alpha}$ globin locus can be grouped by similarity.} 
 A clustering analysis gives a dendrogram (left) which indicates how similar or different the conformations are. Conformations fall into four main representative structures depending on the pattern of contacts they exhibit (see \supmethods{}). Contact maps for each representative structure are shown (centre; the region shown is indicated by the green line in \fref{1C}), as is a schematic of each representative structure (right). The proportion of simulated conformations adopting a given structure gives a prediction of the frequency with which that structure will occur in a population of cells.}
\end{figure}

As anticipated, one of the main strengths of our approach is that it naturally outputs information on each member of the population of chromatin conformations (these can be thought of as representing different cells, or the same cell at different times), which we can then further interrogate. A clustering analysis (i.e., grouping the conformations by similarity; see \supmethods{} for details) of 1000 simulated conformations reveals that the locus folds into four main representative structures (\fref{2}). The main distinction between these structures is whether a single bridging-induced globular domain forms (of size $\sim$70~kbp), or whether it breaks into two smaller microdomains, one containing around 40~kbp, and the other one around 25~kbp. The size of these globular microdomains does not exceed 100~kbp, so these are much smaller than TADs (the median size of a TAD is~1 Mbp~\cite{hicdixon}); interestingly, though, their size is comparable to that of the sub-TAD domains observed within active regions~\cite{hic14}, and also to that of the so-called supercoiling domains recently found in mammalian cells~\cite{nick}.

In the most common representative structure, which accounts for 53\% of the total observed conformations for the locus, there is a single globular domain containing the promoters of the globin genes, the promoters of the two neighbouring genes \textit{Mpg} and \textit{Nprl3}, and all five known regulatory elements. A similar representative structure, which accounts for 6\% of conformations, also has a single globular domain, but the region which contains the \textit{Nprl3} promoter is in a loop outside the globule. A third representative structure accounts for 14\% of the conformations: here two globular microdomains form, where the $\alpha$ genes interact with only the two genomically closest regulatory elements. The fourth structure, which is adopted by about 25\% of the conformations, has again two microdomains, but their composition is different: now the $\alpha$ genes are no longer in the same microdomain as the regulatory elements. We expect that these genes should be transcriptionally inactive when the locus adopts this structure. Finally, there are a small number ($\sim1\%$) of conformations which do not fit into any of these four clusters.  It is also interesting to note that the $\zeta$ gene and \textit{Mpg} seldom interact with the elements (these genes are not widely expressed in adult erythroid cells). The arrangement within the domains can be further probed by looking at which promoters are directly interacting with the different regulatory elements in each conformation (see \sfref{3}). We find, for example, that one or more of the $\alpha$ promoters interacts with one or more of the elements in 65\% of conformations, and that \textit{Hba-a1} interacts with the elements in 53\% of conformations whereas \textit{Hba-a2} interacts in only 41\%. This is qualitatively consistent with experiments in which mRNA expression from the two $\alpha$ globin paralogues was measured independently (on the basis of 3$'$ sequence divergence), which showed that the gene situated linearly closer to the enhancer elements, \textit{Hba-a1}, is always expressed at a higher level~\cite{Higgs2008}.

\begin{figure}[th!]
\includegraphics[width=0.475\textwidth]{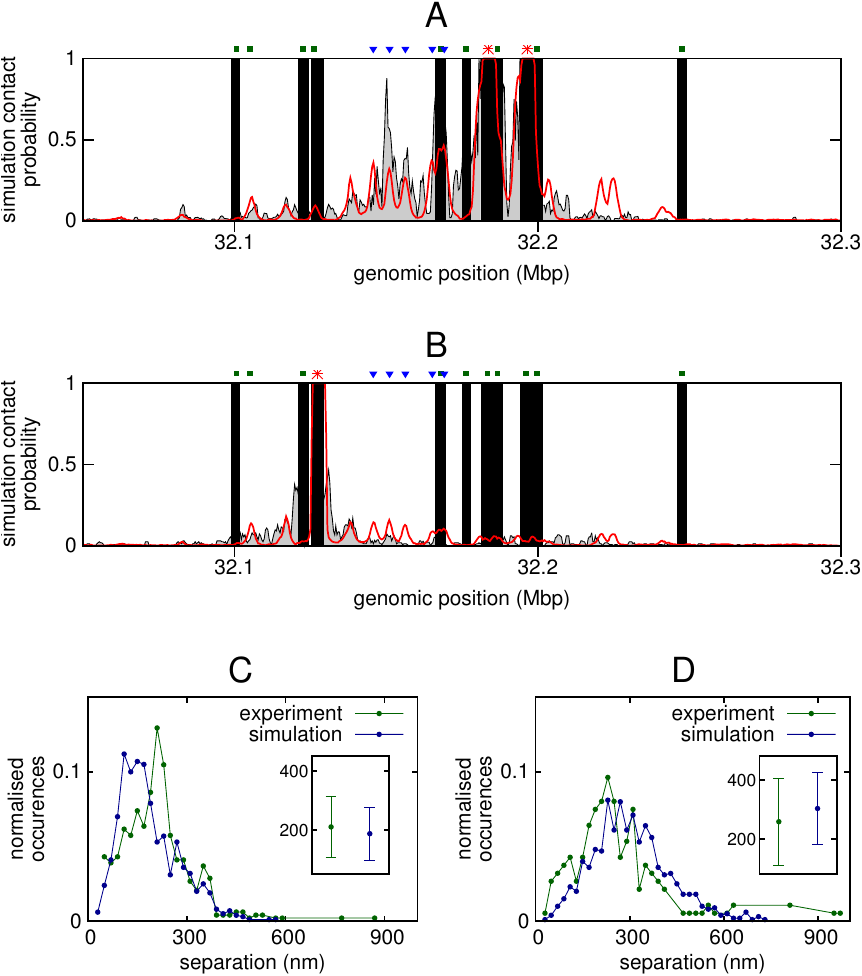}  
\caption{\textbf{Simulations compare favourably with experimental data.} 
(A) Plot showing the contacts made with the promoters of the two $\alpha$ globin genes (locations indicated by red asterisks; the positions of the regulatory elements and other gene promoters are also indicated). Simulation results (red) are shown alongside Capture-C data (grey); in both cases the plots show the contacts to both genes combined (since each copy of the gene has the same sequence it is impossible to separate these in the experiment). Black bars indicate regions where there is no contact data (i.e. between captured regions; see \supmethods{} and Ref.~\cite{Hughes}). Since Capture-C data only gives relative contact strength, the height of the experimental data has been scaled so as to best fit the simulation results (see \supmethods{}). (B) As in A, but now showing the contacts made with the \textit{Mpg} promoter (position indicated by red asterisk). Although \textit{Mpg} is roughly the same genomic distance away from the regulatory elements as the $\alpha$ globin genes, it interacts with them less frequently. 
(C) Plot showing the distribution of the 3-D separation of the $\alpha$ globin promoters and the probe pE located at the regulatory elements R1-3. Simulations are compared with FISH measurements (see \matmeth{} and \sfref{5}) performed on mature erythroblasts 30~hours after differentiation, when the globin genes are maximally expressed. The inset shows the mean and standard deviation for each case. 
(D) As in H, but the separation of the $\alpha$ promoters and a downstream control probe p58 located within the \textit{Sh3pxd2b} gene.}
\end{figure}

Importantly, we can also compare the interactions predicted by our simulations with recent high-resolution Capture-C data~\cite{Hughes} which mapped the chromosomal contacts within a number of \textit{cis}-regulatory landscapes in mouse erythroblasts (see \supmethods{}). Specifically, \fref{3A} compares Capture-C and {\it in silico} patterns of contacts with the promoters of the two $\alpha$ globin paralogues (which cannot be separated in the experimental data as they share the same sequence). \fref{3B} shows a similar plot for the \textit{Mpg} promoter. The results show that, remarkably, {\it with the sole input} of the ChIP-seq and DNase-seq data giving the locations of the protein binding sites, we can reproduce to a good accuracy the Capture-C profiles. In particular, we reproduce the contacts between the $\alpha$ promoters and the five known regulatory elements; we also reproduce the fact that there is some interaction between the regulatory elements and the \textit{Nprl3} promoter (see \sfref{4}), but far fewer interactions with the \textit{Mpg} promoter, despite the fact that this gene is a similar genomic distance away from the elements as the $\alpha$ genes. 

To further assess the level to which the population of locus conformations predicted by our model gives a faithful representation of the organisation of the $\alpha$ globin locus in real cells, we performed FISH experiments (see \matmeth{}) to obtain distributions of the separations of probes at different positions across the locus. These measurements also allow us to parametrise the physical size of the 400~bp simulation beads by fitting the means of each distribution (see \matmeth{} and \sfref{5}); this is the only fitted parameter in our model, and the fit yields a size of 15.8~nm, which is reasonable given that 400~bp corresponds to two nucleosomes. Plotting the experimental and simulation separation distributions on the same axes (\frefs{3C-D}, and \sfrefs{5}D-G) reveals that once more the simulations give an accurate prediction of the structure of the locus; for example the separation of the $\alpha$ promoters and pE at the regulatory elements R1-3 shows a narrow distribution peaked about a mean value of $\sim200$~nm, whereas the separation of the promoters and a probe p58 at roughly the same genomic distance, but telomeric to the locus, shows a much broader distribution with a mean closer to 300~nm.

We can also define a quantitative score $\mathcal{Q}$, taking values between 0 and 1, which indicates how well our simulations predict the experimental Capture-C interaction profiles (see \supmethods{} for details). By combining Capture-C data from a number of promoters across the locus, we can obtain a mean $\mathcal{Q}$ value along with a standard error (\sfref{6}). This allows us to compare results from different model set-ups. Specifically, we examined the effect on the experiment-simulation comparison scores of changes in: (i) chromatin stiffness; (ii) number of bridges; and (iii) level of coarse-graining (see \supmethods{} and \sfref{6}). For the first two cases we find only a modest effect on the $\mathcal{Q}$-score for the simulated configurations (\sfref{6}); if we decrease the resolution of our model by changing the coarse-graining, then this performs less well. Interestingly the representative structures found from the clustering analysis of the population of conformations found {\it in silico} are always the same. What changes in some cases is the proportion of conformations which adopt each representative structure. In the model where the chromatin was stiffer, the globular microdomain structure containing all of the regulatory elements occurred less often, whereas the structure where the \textit{Nprl3} promoter loops out was more likely; this is because holding the \textit{Nprl3} promoter in the microdomain requires bending of the chromatin fibre, which is disfavoured when this is stiff.  Also, when we examined the effect of changing the number of protein complexes in the simulations, we found that, as more proteins are introduced, there is a greater likelihood that the locus adopts a structure with two globular microdomains; this is because forming more protein bridges between chromatin binding regions, while being energetically favourable, leads to the formation of more loops whose entropic cost increases non-linearly with the number of loops~\cite{enzoJSTAT}.

\subsection*{Chromatin folding of the $\boldsymbol{\beta}$ globin locus}

\begin{figure*}[th!]
\includegraphics{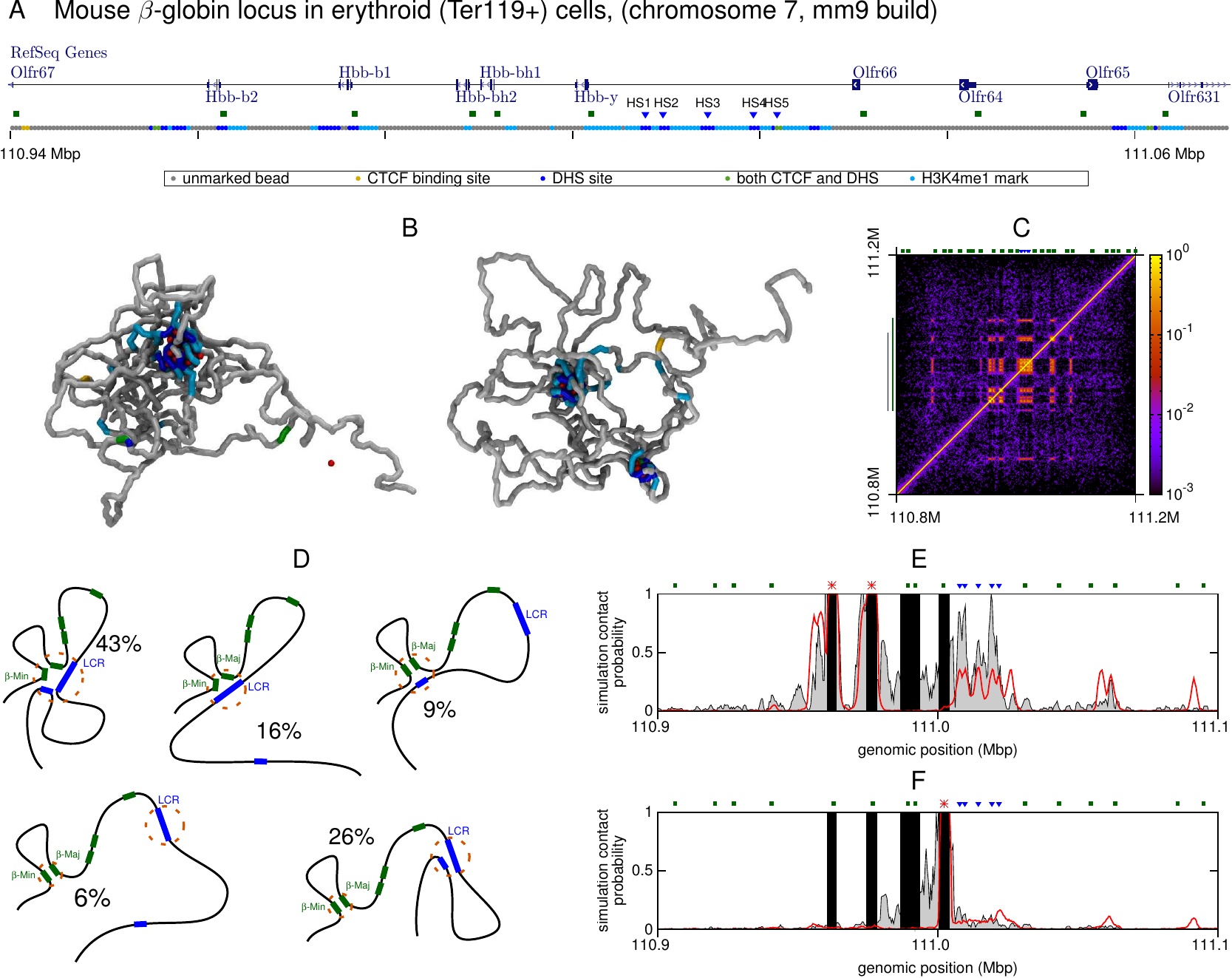}
\caption{\textbf{\textit{Cis}-interactions of the $\boldsymbol{\beta}$ globin locus.} 
(A) Browser view showing genes in the vicinity of the $\beta$ globin locus, alongside a schematic indicating the coarse-graining used in the simulations. A 130~kbp section of the 400~kbp chromatin fragment which was simulated is shown. The positions of the known regulatory elements within the LCR are indicated with blue triangles, and promoters with green squares. 
(B) Example simulated configurations of the locus. CTCF proteins (green) and DHS binding proteins (red) are shown; the chromosome fragment is coloured as in A. 
(C) Contact map showing the frequency of contacts between each chromatin bead in 500 simulated configurations. The colour bar shows a logarithmic scale. The blue line to the left indicates the region which is shown in A; the green line indicates the region which is used in the clustering analysis. 
(D) As in \fref{2}, clustering analysis allows conformations to be grouped by their structural features. Schematics of the representative structures are shown, with the \% of conformations in which they occur; a dendrogram, and contact maps for each representative structure are shown in \sfref{8}. 
(E) Plot showing the contacts made with the promoters of the two $\beta$ genes (locations indicated by red asterisks; the positions of the regulatory elements and gene promoters are indicated). Simulation results (red) are shown alongside Capture-C data (grey); both cases show the contacts to both genes combined (since each copy of the gene has the same sequence it is impossible to separate these in the experiment). Black bars indicate regions where there is no contact data (see Ref.~\cite{Hughes} and \supmethods{}). (F) Similar plot showing the contacts made with the \textit{Hbb-y} gene (position indicated by red asterisk).
}
\end{figure*}

We have also applied our chromosome-and-bridges model to the mouse $\beta$ globin locus (chr7:110800000-111200000, mm9 build; \fref{4}, \sfref{7}, and \sfref{8}). This locus contains five globin genes: the $\epsilon$y gene, $\beta$h1 and 2, and two $\beta$ globin genes $\beta$-Major and $\beta$-Minor. The expression of each gene depends on the stage of development (the $\epsilon$y and $\beta$h1 genes are predominantly expressed in embryos, while the $\beta$ genes take over in adults), and is controlled by interactions with a series of DHSs in a region known as the locus control region (LCR)~\cite{Trimborn1999,Palstra2003}. Unlike the $\alpha$ globin locus, the $\beta$ globin genes are surrounded on either side by a condensed chromatin region, containing genes which are not expressed in erythroid cells. As with the $\alpha$ globin case, we use ChIP-seq and DNase-seq data to label a bead-and-spring polymer which represents the gene locus (see \fref{4A}, and \sfref{7}). A clustering analysis of a population of 500 simulated conformations reveals that the most abundant representative structure of the $\beta$ globin locus (43\% of the total conformations, see schematics in \fref{4C} and dendrogram in \sfref{8}), features a single globular domain, where the $\beta$ Major and Minor promoters co-localised with the five regulatory elements in the LCR, and with a CTCF site on the telomeric side near the \textit{Olfr65} gene. A further 16\% of conformations adopt a similar representative structure, but the promoters interact only with the LCR. We also note that when the locus adopts these structures, there is an interaction between the CTCF sites in the LCR and the one on the centromeric side of the $\beta$ genes near the \textit{Olfr67} gene (these contacts are just visible on the left and bottom edges of the top two contact maps in \sfref{8A}) which has previously been observed in both definitive erythroblasts and erythroid progenitors, but is absent in non-erythroid tissue~\cite{Tolhuis2002,Palstra2003}. This is consistent with the hypothesis that CTCF mediated loops in progenitors hold the locus in a structure poised to facilitate $\beta$ globin expression upon differentiation~\cite{Palstra2003} (though see below). A third representative structure, which accounts for 9\% of the simulated conformations, has the $\beta$ promoters interacting only with the DHS near \textit{Olfr65}. The Capture-C data, along with previous work~\cite{Tolhuis2002,Palstra2003}, confirms the prediction that this site (usually denoted HS-60) interacts with the $\beta$ globin promoters; indeed it has been previously shown that there are interactions between all hypersensitive sites in the locus~\cite{Tolhuis2002} and the pair of sites HS-60/-62 are normally taken to demarcate the boundary of the locus. Whether this particular DHS (HS-60) has enhancer properties remains unclear, however it binds Scl/Tal1  (a transcription factor thought to play a key role in hematopoietic differentiation~\cite{Kassouf2010}), is near to a CTCF binding site (HS-62), and is within a region marked by monomethylation of histone H3 Lys4, which is normally associated with enhancers. In the remaining 32\% of the conformations (bottom two schematics in \fref{4D}), the $\beta$ globin promoters are still together, but do not interact with the hypersensitive sites (\sfref{8A}). 

We note that the microdomains which form in each type of the five representative structures have more ``looped out'' regions (consistent with conclusions from 3C experiments in Ref.~\cite{Tolhuis2002}) than in the $\alpha$ globin locus (compare contact maps in \freftwo{1C}{2} with \fref{4C} and \sfref{8A} -- more gaps are seen between the blocks of highly probable interactions in the $\beta$ globin case). This indication that the $\beta$ globin locus is less compact than the $\alpha$ globin case is borne out in measurements of the overall 3-D size of the simulated loci (see distributions of the radius of gyration of the polymer in \sfref{9}G compared to the $\alpha$ globin case in \fref{7G}).

As in the case of the $\alpha$ globin locus, our simulations predict contact patterns which are in good agreement with Capture-C data, both for the $\beta$ Major and Minor gene promoters (\fref{4E}) and for the \textit{Hbb-y} promoter (\fref{4F}). This demonstrates that our model is not gene-specific, but can be applied, in principle, genome-wide, at least to active regions; the two bridges which we model, CTCF and DHS binding proteins, are indeed found in most euchromatic, open chromatin, regions. Given its relatively low computational cost (harvesting 500 conformations for a 400 kbp chromosome region at a 400~bp resolution can be done in about a day with a multi-core machine, see \supmethods{}), we expect this modelling to be useful in predicting the overall folding of previously uncharacterised active chromosomal loci -- the knowledge of the predicted population of 3-D structures can then direct further high-resolution Hi-C, Capture-C or fluorescence hybridisation experiments (as in \freftwo{3}{4E-F}) to characterise that region more accurately.

\subsection*{The model accurately reproduces differences in locus folding across cell types}

\begin{figure*}[th!]
\includegraphics{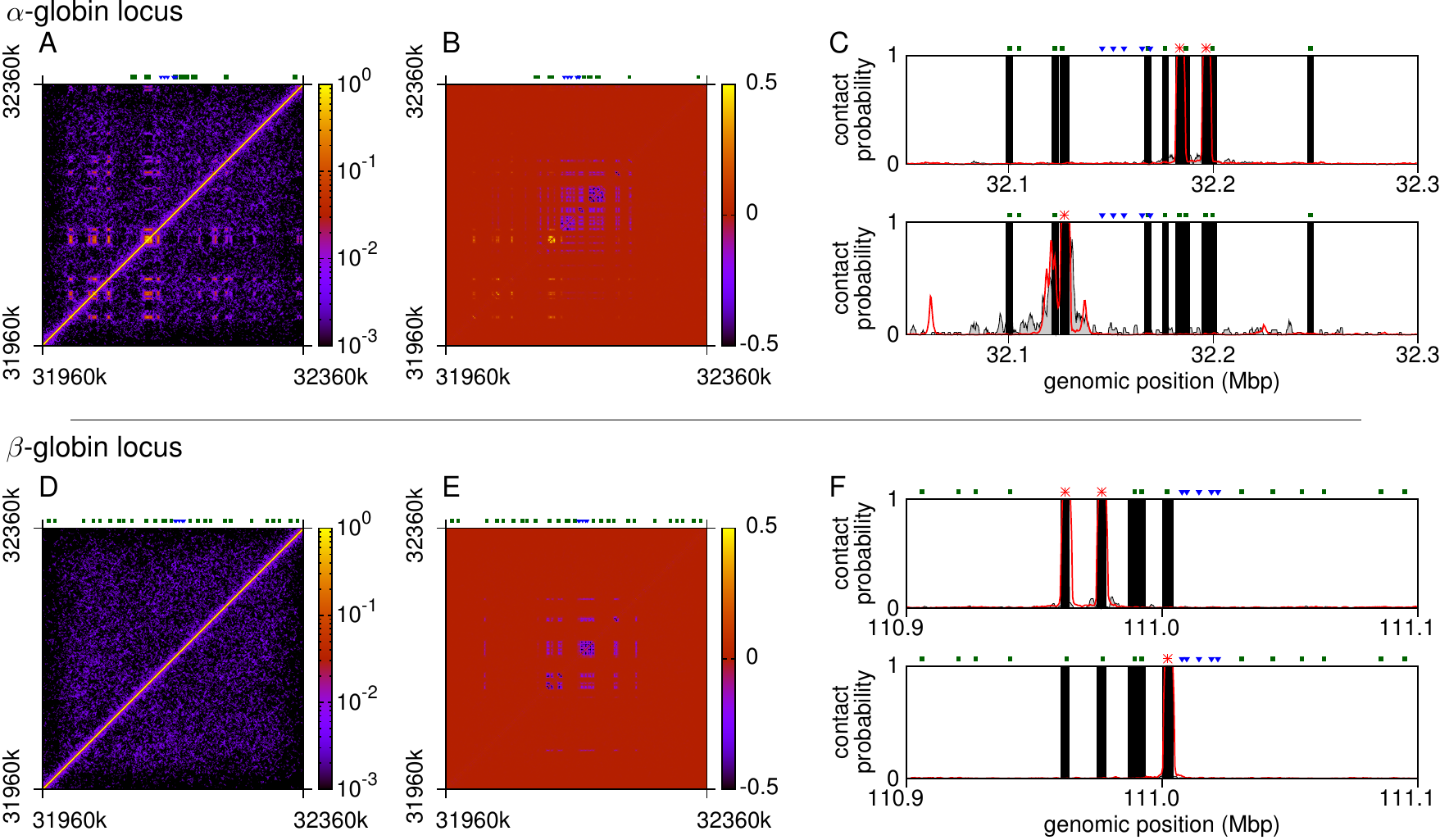} 
\caption{\textbf{Simulations show changes in locus organisation across cell types.} (A) Contact map for 500 conformations for the $\alpha$ globin locus in mouse embryonic stem cells (mES). Simulations are performed as in \fref{1}, but using mES ChIP-seq and DNase-seq data, as shown in \sfref{9}. (B) Difference between the contact maps in panel A and \fref{1C}. Blue regions indicate contacts which were present in erythroblasts, but not mES, and yellow indicates contacts present in mES but not erythroblasts. (C) Plots comparing simulations and Capture-C data for mouse embryonic stem cells (data from Ref.~\cite{Hughes}). (D)-(F) Similar plots but for the $\beta$ globin locus.}\label{stemcells}
\end{figure*}

Importantly, because data showing protein binding, hypersensitive sites and histone modifications are available for different cell types, we can also predict changes in the three-dimensional organisation of a chromosomal region across cell types or at different times in development. We show in \fref{5} how the folding of the globin loci differs in mouse embryonic stem cells (where the globin genes are inactive) with respect to the organisation predicted for erythroblasts. The bioinformatic data used to inform our modelling for stem cells are given in \sfref{10}.

\fref{5A} shows the contact map predicted from simulations of the $\alpha$ globin locus. Our model predicts that in ES cells the contacts are much sparser than in erythroblasts, that the bridging-induced domain around the $\alpha$ globin gene is lost (\fref{5B}), and that no interactions with the regulatory elements are observed; the same is true of the neighbouring \textit{Mpg} promoter. Once again, the contacts observed {\it in silico} reproduce the experimental ones (\frefs{5C}), with some minor inaccuracies for \textit{Mpg} (which likely originate from our approximation that all DHSs are the same in regards to bridge formation, but nevertheless highlight the principle that the locus can adopt a completely different shape in a different cell type). When repeating the analysis for the $\beta$ globin locus we find that the loss of non-local contacts is even more dramatic (\frefs{5D-E}), and the agreement with the data even more remarkable (\frefs{5F}), with all non-local (i.e. off diagonal) interactions being absent.

\begin{figure*}[th!]
\includegraphics{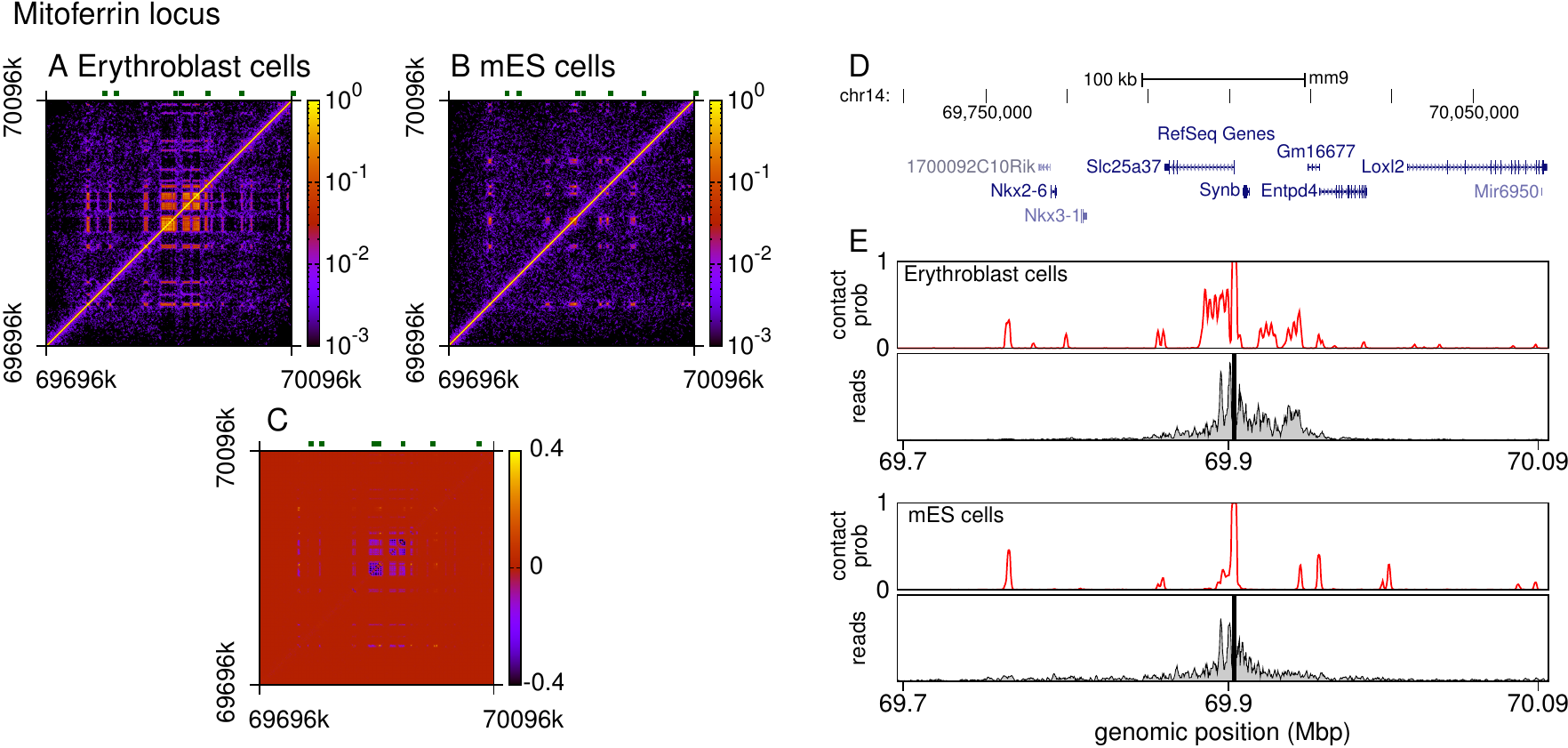} 
\caption{\textbf{Simulations also correctly predict looping for a less studied locus.} Simulations of the \textit{Slc25a37} gene (Mitoferrin1) were performed for mouse erythroblasts and embryonic stem cell, using a similar input data as for the globin loci (DNase-seq, and ChIP-seq for CTCF and the H3K4me1 histone modification). (A) Contact map from the simulations of erythroblasts showing the frequency of contacts between each chromatin bead in 500 simulated configurations. (B) Similar contact map for the same locus in mouse embryonic stem cells. (C) Difference between the contact maps in panels A and B. Blue regions indicate contacts which were present in erythroblasts, but not mES, and yellow indicates contacts present in mES but not erythroblasts. (D) Browser view showing the genes across the 400~kb simulated region. (E) Plots showing the interaction profiles for the \textit{Slc25a37} promoter in each cell type, comparing simulation results (upper panels) with new Capture-C data (lower panels). Note that the genomic coordinates are aligned with the browser view in D.}
\end{figure*}

To further demonstrate the wide applicability of the model, we also perform a set of simulations for a region surrounding the \textit{Slc25a37} (Mitoferrin1) gene in both mouse erythroblasts and embryonic stem cells. This gene encodes a mitochondrial protein essential for iron import into mitochondria, however much less is known about this locus than about the $\alpha$ or $\beta$ globin, and so our results represent a true prediction of its folding. The input data used was similar to that of the globin loci, and are given in \sfref{11}. As shown in \fref{6} the simulations predict that in the erythroid cells (where the gene is active) the locus forms a compact domain around \textit{Slc25a37} and \textit{Entpd4}; the \textit{Slc25a37} promoter interacts strongly across the \textit{Slc25a37} gene, but also with two distinct regions between the nearby \textit{Synb} and \textit{Gm16677} genes (\fref{6E}, top panel). These are enriched for mono-methylation of Lysine 4 of Histone H3 (see \sfref{11D}), suggesting that sites within these regions have enhancer activity (as was also proposed in Ref.~\cite{Amigo2011}). In order to test these predictions we compare with new Capture C experiments (performed as detailed in Ref.~\cite{Hughes}). As before, our very simple model gives a remarkable agreement with the data: strong interaction with the putative enhancer regions is observed in the erythroid, but not the stem cells. Some longer distance interactions which are predicted in both cell types are not found in the experimental data; these errors are due to our approximation that bridges can form between any DNase hypersensitive sites, and the agreement would likely be improved with a different choice of input data (e.g. using TFs involved in regulation of this gene). 

\subsection*{The typical 3-D structures of the globin loci are preserved in CTCF or other TF knock-outs}

Another strength of our approach is that it is easy to alter the protein binding profiles in our simulations to investigate e.g. genome modifications or protein knock-outs etc., and predict the consequences of these for the 3-D organisation \textit{in vivo}. For example, we can switch off interactions with the hypersensitive sites, and only include the CTCF bridges in the simulation, or simulate a CTCF knock-out by switching off interactions with the CTCF sites and any hypersensitive sites where only CTCF binds (i.e. DHSs which bind CTCF, but none of the other TFs implicated in globin regulation).

\begin{figure*}[th!]
\includegraphics{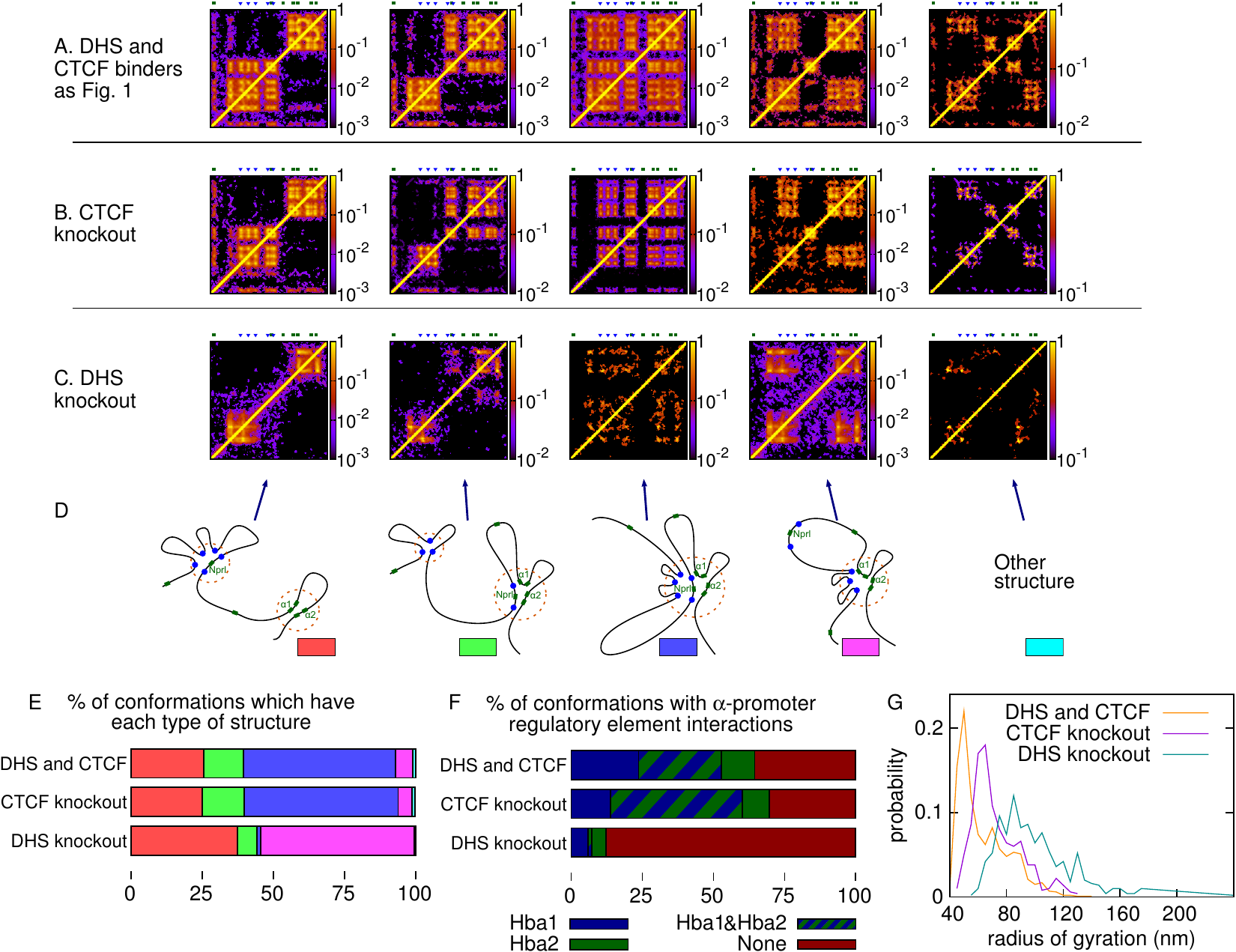} 
\caption{\textbf{Simulations predict the effect of protein knock-outs in the $\boldsymbol{\alpha}$ globin locus.} Plots showing the effect of a CTCF knock-out, and a ``DHS knock-out'' (equivalent to knocking out all protein complexes involved in looping the $\alpha$ globin locus \textit{except} CTCF). 
(A)-(C) Contact maps showing the interactions between different chromosomal locations for conformations within each group identified by clustering analysis. Maps from three sets of simulations are shown; the positions of the known regulatory elements and gene promoters are indicated above each plot. 
(D) Schematics showing the structure of the locus within each group. 
(E) Plot showing the percentage of conformations which belong to each group identified by the clustering analysis. The colour key is given in D. 
(F) Plot showing in what percentage of conformations the two $\alpha$ globin gene promoters are interacting with one or more of the known regulatory elements.
(G) Plot showing the distribution of the radius of gyration of the locus across the simulated conformations. The radius of gyration is defined as $R_g^2=(1/N) \sum_{i=1}^N (\mathbf{r}_i-\mathbf{\bar{r}})^2$, where $\mathbf{r}_i$ is the position of the $i$th chromatin bead in the polymer, and $\mathbf{\bar{r}}$ is the mean position of all $N$ chromatin beads.}
\end{figure*}

In the case of the $\alpha$ globin locus we find that, surprisingly, for both the CTCF and DHS knock-outs the same folded structures can still form (\frefs{7A-D}). For the CTCF knock-out, the relative proportions of each structure found in the clustering analysis remain largely unchanged (\fref{7E}): the most common one is again the single globular domain containing the $\alpha$ promoters and all regulatory elements. If we assume that the level of $\alpha$ globin expression correlates with the fraction of conformations in which one or more of the $\alpha$ promoters is interacting with one or more of the regulatory elements, then this expression level also remains largely unchanged (the genes are active in 65-70\% of conformations, see \fref{7F}). For the DHS knock-out on the other hand, the number of conformations showing regulatory element interactions drops to less than 20\%. There is also a change in the proportions of the different groups found by the clustering analysis, with the structure in which the \textit{Nprl3} promoter loops out of a single domain becoming most common. Nevertheless, it is remarkable that despite loss of binding at the regulatory elements (which presumably reduces $\alpha$ globin expression), the CTCF sites near the \textit{Hbq1} and \textit{Hbq2} promoters, and within the introns of the \textit{Nprl3} gene (green and yellow in \fref{1A}) are sufficient to allow the locus to fold into the same representative structures. We can also measure the effect on the overall size of the domain by calculating the radius of gyration of the polymer; \fref{7G} shows the distribution for each of the {\it in silico} knock-outs. We see that loss of protein binding generally leads to an expansion of the locus, with the DHS knock-out having more effect than the CTCF case.

A similar scenario applies to CTCF and DHS knock-outs in the $\beta$ globin locus (\sfref{9}). Here, however the contact map for each of the groups identified by the clustering analysis (\sfrefs{9}A-C) shows some subtle differences between the knock-outs. Again the CTCF knock-out appears to have little effect, leading to only small changes in the fraction of simulations adopting each structure or the contacts between the $\beta$ promoters and the LCR. The DHS knock-out leads to a notable reduction in the promoter-LCR interactions, and a reduction in the number of conformations adopting the structure where the $\beta$ promoters interact with the hypersensitive site near the \textit{Olfr65} gene. This locus also expands upon protein knock-outs, albeit to a lesser extent than the $\alpha$ globin case; this is probably due to the $\beta$ globin locus being less compact initially.

Given the suggestion that CTCF proteins play a key role in genome organisation, it might seem surprising that the knock-out simulation shows a relatively minor change in the folding structures and promoter-enhancer interaction in both globin loci. However, CTCF is known to have a variety of different functions, for instance it acts as a barrier against the spreading of repressive heterochromatin, or as an insulator, preventing interactions with other nearby chromosome regions~\cite{Holwerda2013}. A recent study suggested that a depletion of CTCF has only a mild effect on the domain organisation of chromosomes as found via Hi-C experiments~\cite{Zuin2014}, and a ChIA-PET analysis of the contacts made between CTCF-bound regions found that, only a fraction of the 40,000 CTCF binding sites are involved in these~\cite{Handoko2011}: presumably, this fact is related to the recently discovered importance of CTCF binding site directionality in loop formation~\cite{hic14,deWit2015,Imakaev2015}.
In the specific case of the $\beta$ globin locus, another recent study found that reducing the abundance of CTCF protein, or disrupting a specific CTCF binding site within the locus in erythroid progenitor cells leads to a loss of chromosome looping; however upon differentiation to mature erythroblasts, these cells are still able to express $\beta$ globin, and fruitful interactions between the promoters and the LCR can still form~\cite{Splinter2006} (i.e. setting up loops in progenitor cells appears not to be necessary). Together this suggests that the globin loci may be examples where CTCF-mediated chromosome loops are not crucial in determining the 3-D organisation, though of course CTCF is likely to have some other function (e.g. protecting other nearby genes from activation) and may still play an important organisational role at a larger scale~\cite{Hou2010}. In our simulations the CTCF bridges certainly do form loops, but in their absence the overall folding patterns can be maintained by the other bridges.

\section*{Discussion}

In this work we have shown that a minimal polymer model informed by large bioinformatic datasets on protein binding can successfully reproduce the pattern of Capture-C contacts observed in the well studied $\alpha$ and $\beta$ globin loci within mouse erythroblasts (a cell type where these genes are highly active), and also within the less understood \textit{Slc25a37} (Mitoferrin1) locus. Our model is built on the hypothesis that there exists architectural protein bridges, which we assume are either CTCF, or generic bridges made up by complexes of transcription factors and other DNA-binding proteins. The only inputs we require are ChIP-seq data for CTCF binding, and the map of DNase1 hypersensitive sites, which we take as a proxy for the location of the binding sites for the generic protein bridges (DHS bridges). Importantly, our approach differs from other recent polymer modelling studies which also have predictive power~\cite{tiana,Bau2011,Lesne2014}, in that it does not rely on fitting to pre-existing 5C or Hi-C data. Due to this feature, it can be applied to relatively poorly characterised loci (e.g., Mitoferrin1, see~\fref{6}), for which only few data exist (e.g., DNase tracks); the model can then be developed when needed as more experimental data become available. 

Our model generates a population of conformations, hence we can predict, for instance, the distribution of distances between selected targets on the globin locus. These results compare very favourably with our FISH measurements, which allow us to estimate the physical size of the beads in our coarse-grained polymer (or equivalently, the DNA packing density in the chromatin fibre in the globin locus; this is the only fitting parameter in our model). The packing we obtain (15.8~nm for 400~bp) is consistent with open chromatin, which is reasonable since the region we focus on is highly active. 

The fact that our model generates a population of conformations, rather than a single average conformation, is important because it gives an estimate of the stochasticity and fluctuations in \textit{in vivo} 3-D organisation. A key result of our model is that the conformations of the loci we studied can be grouped into a handful of representative structures, which account for different fractions of the whole population. In both the $\alpha$ and $\beta$ globin loci, the analysis suggests that there is a split in these structures between two main types: those in which there is a single globular domain which includes the active genes together with their regulatory elements, and those where the globule splits into two microdomains. The single globule structures are favoured by bridging, while the competing structure requires less bending and looping, and costs less entropy. (This is because there are more ways to place two microdomains in space than there are for a single one, and also because the entropy of forming $n$ loops in the same place scales non-linearly with $n$~\cite{enzoJSTAT}). There is a subtle balance between these contributions, which are both of the order of a few $k_BT$, therefore both structures coexist in the population. A consequence of this is that the globin loci are naturally poised close to a transition between two different 3-D folding phenotypes; because the competition between bridging and entropy is likely to be a generic feature, we suggest that the plasticity associated with this balance between competing effects may be an underlying principle in the organisation of active regions genome-wide. This suggests that the cell could tip the balance one way or another by changing the abundance or specificity of bridges, or the properties of the fibre (e.g. by histone modification or chromatin remodelling). 

In future work it will be interesting to compare these predictions with experimentally determined chromatin dynamics through cell differentiation, for example examining the $\alpha$ globin genes using techniques that permit imaging of the locus during erythroid differentiation in live cells. Another application of the work might be to provide some explanation of how the \textit{Hba-x} gene is silenced in adult erythroblasts: in all of our predicted conformations it does not contact the known enhancer elements nor the surrounding gene promoters. It may also be informative to repeat the modelling for primitive erythroblasts, when sufficient protein binding and DNase hypersensitive data becomes available for that cell type.
 
As we have seen, our model can be further exploited to predict the organisational consequence of the knock-out of proteins such as CTCF (or our generic DHS bridge). Similarly, one can perform an {\it in silico} experiment which follows the consequences of modifying some genomic region within a locus. An intriguing example is the deletion of the R2 (HS-26) hypersensitive site in the $\alpha$ globin locus, which has been shown experimentally to result in a 50\% reduction of $\alpha$ globin RNA levels~\cite{Anguita2002} (a much milder phenotype than the severe $\alpha$ thalassemia which results from a deletion of the equivalent HS-40 element in humans~\cite{Vernimmen2009}). Removing the R2 site in our simulation only leads to a $\sim3\%$ reduction in the number of conformations where the $\alpha$ promoters interact with the remaining regulatory elements. We can make our model more complex by replacing DHS binding proteins with bridges which bind to specific TF binding sites. For instance, GATA1 and Klf1 are a minimal set of TFs (see \sfref{2}) which can interact to form bridges between the $\alpha$ globin promoters and the regulatory elements, and which can discriminate between the different elements (i.e. GATA1 binds to R1-4 only, whereas Klf1 binds to R2, and the $\alpha$ promoters only). Thus we use a model with three protein species, binding strongly to GATA1, Klf1, and CTCF sites respectively (no longer considering hypersensitive sites), and weakly to H3K4me1 modified regions (using ChIP-seq data as shown in \sfref{2}), and repeat the \textit{in silico} R2 knock-out experiment (see \sfref{12}). Quite remarkably, in a ``wild type'' simulation, this more detailed model reproduces the differences in peak heights for interactions between the $\alpha$ promoters and elements R1-3 as shown in the Capture-C data (i.e. there is a higher probability of interaction with R2 than R1 and R3; \sfref{12}A). For the R2 knock-out case, the three bridge model shows a $\sim20\%$ reduction in the number of conformations where the $\alpha$ promoters interact with the remaining regulatory elements (much closer to what might be expected given the experimentally observed effect on $\alpha$ globin RNA levels).  Therefore, our approach can be generalised to accommodate more biological detail in a modular fashion, in the cases where this detail is known.  

We anticipate that the main application of our {\it in silico} chromosome folding model will be to investigate regions of mammalian and other eukaryotic genomes which are currently poorly characterised. The approach relies only on DNase hypersensitivity and protein binding data, which are available genome-wide for many organisms and cell types. Our technique is fast and inexpensive, so that it can be used to predict the organisation of a large number of wild-type and modified genomic loci prior to, for example, a combination of detailed Capture-C, 5C or FISH experiments, directing focus to those regions whose predicted structure was deemed to be of particular interest. The ease with which genome modifications can be incorporated makes it highly applicable for investigation of the effect on 3-D chromatin structure of, for example single nucleotide polymorphisms at enhancers, which have been implicated in many diseases.

In the present work we have focussed on looping interactions within a gene locus, at a sub-TAD length scale. Polymer models, and the principal of protein bridges driving chromatin conformations, can easily be adapted to treat larger looping and organisation at the chromosome and genome scale, and this will be the subject of a future study.

\section*{Methods}

\subsection*{Polymer model and simulation scheme} The chromatin fibre is modelled as a simple coarse-grained bead-and-spring polymer, where each bead represents 400~bp of DNA, or roughly two nucleosomes. The positions of the beads are updated via a molecular dynamics scheme (Langevin dynamics) using the LAMMPS (Large-scale Atomic/Molecular Massively Parallel Simulator)~\cite{lammps} software. Pairs of beads adjacent along the polymer back-bone interact via finitely extensible non-linear elastic (FENE) springs, and the polymer is afforded a bending stiffness via a cosine interaction between triplets of adjacent beads. We choose parameters such that the persistence length is 4 beads, which is reasonable for euchromatin~\cite{Langowski2006}. The beads also interact with each other via a Weeks-Chandler-Anderson potential, meaning they cannot overlap. Protein complexes are modelled as single spheres which interact with each other also via a Weeks-Chandler-Anderson potential (i.e. they have a steric interaction only). Each chromatin bead represents a region of the chromosome locus of interest, and is labelled as binding or not for the various protein species according to the input data. Proteins interact with chromatin beads labelled as binding via a shifted, truncated Lennard-Jones interaction which has short-range repulsive and longer-range attractive parts; they interact with non-binding chromatin beads again via the Weeks-Chandler-Anderson potential. Full details of all interaction potentials are given in \supmethods{}, and parameter values in \stref{1}. As an input to the model we use ChIP-seq and DNase-seq data (see Additional files 2, 7 and 10: Figures S2, S7 and S10; data from Refs.~\cite{Hughes,Marques2013,Kassouf2010,Tallack2010,encode} as indicated in figure captions) to identify protein binding sites in the chromosome region of interest.  Full details of the bioinformatics data analysis are given in \supmethods{}.

\subsection*{Capture-C data} The Capture-C data shown in \frefs{3-5} and \sfref{4} were previously published in Ref.~\cite{Hughes}. For \fref{6} new Capture-C experiments were performed using the same methods and cell lines as Ref.~\cite{Hughes}. Full details of how the data were processed so as to compare with the simulation results are given in \supmethods{}.

\subsection*{Fluorescence in-situ hybridization data} \frefs{3C-D} and \sfrefs{5}C-G show distributions of the separation of probe pairs at different locations in the $\alpha$ globin locus in mouse erythroblasts, where the $\alpha$ genes are active. Genomic locations of the probes are given in \sfref{5}A. Probes were constructed in the pBS plasmid by subcloning regions from mouse BACRP23-469I8  and BACRP24-278E18 (obtained from CHORI) by $\lambda$-Red mediated recombination using oligonucleotide sequences shown in \stref{2} (below). Recombineering was carried out mixing 50~$\mu$l of cells with 150~ng to 300~ng of purified DNA in a 0.1~cm wide cuvette using a Bio-Rad gene pulser set at 1.8~kV. Immediately after electroporation, 1~ml of SOC media was added, and cells were further grown at 37$^\circ$C for 1~hour before being plated on selective agar media containing 100~$\mu$g/ml ampicillin.

\textit{In vitro} cultured mouse foetal liver cells (expressing $\alpha$ and $\beta$ globin genes) were settled on poly-l-lysine coated coverslips, fixed with 4\% paraformaldehyde in 0.25~M HEPES and permeabilised with 0.2\% Triton-X 100. FISH was performed using 7~kbp plasmid FISH probes, labelled with either Cy3-dCTP (GE Healthcare Life Sciences) or digoxygenin 11-dUTP (Roche Life Science). The genomic locations of the FISH probes is shown in \sfref{5}A. Probes were hybridised in pairs (as in \sfrefs{5}B,D-G). Following hybridisation and detection using sheep anti-digoxygenin FITC (Roche Life Sciences) and rabbit anti-sheep FITC (Vector Laboratories), nuclei were imaged on a Deltavision Elite (GE Healthcare Life Sciences) using x100 super-plan apochromat oil 1.4~N.A. objective (Olympus) with a $z$-step size of 200~nm. Images were restored by deconvolution using Huygens Professional software (Scientific Volume Imaging). Probe signal pairs were analysed using a specifically designed Fiji algorithm that measures the 3-D euclidean distance (in microns) between thresholded signal centroids. Each measurement was adjusted to account for chromatic shift by using a displacement vector calculated from 0.1~$\mu$m Tetraspeck\texttrademark microspheres (Life Technologies) collected using the same imaging parameters as in the experiments.

We can parametrise the physical size of the chromatin beads in our simulations by fitting to the mean separation of each pair of probes as measured in the experiment. \sfref{5}B shows a scatter plot of mean values from each pair of probes, with error bars showing the standard error in the mean; we use a linear least-squares fit weighted using the experimental error in the mean to estimate the bead diameter as 15.8~nm. Since we fit to the mean for all probe pairs, the quality of the predicted distributions can still be assessed by comparing the simulation and experiment for each individually.

\section*{Availability of supporting data}

The data sets supporting the results of this article are available in the Edinburgh DataShare repository, [http://dx.doi.org/10.7488/ds/1306], including the new experimental data, simulation output data, simulation input data and scripts. Simulations were performed using the LAMMPS Molecular Dynamics Simulator~\cite{lammps}, which is an open-source code [http://lammps.sandia.gov]. Previously published data used in the work are available at the Gene Expression Omnibus database under accession numbers GSE49460 (DNase-seq, H3K4me1 and H4K4me3 ChIP-seq for Ter119+ cells), GSE21877 (Scl/TAL1 ChIP-seq for Ter119+ cells), GSE20478 (Klf1 ChIP-seq for Ter119+ cells), GSE47492 (CTCF, GATA1, and Nfe2 ChIP-seq for Ter119+ cells), GSE47758 (capture c data for the $\alpha$ and $\beta$ globin loci in Ter119+ and mES cells), and GSE67959 (capture c data for mitoferrin1 in Ter119+ and mES cells). Other datasets used were obtained from the ENCODE project (UCSC Accession wgEncodeEM001703 for CTCF ChIP-seq in mouse ES cells; wgEncodeEM003417 for DNase-seq in mouse ES cells; and wgEncodeEM001681 for H3K4me1 in mouse ES cells).

\section*{Competing interests}
The authors declare that they have no competing interests.

\section*{Author's contributions}
CAB carried out simulation work and bioinformatics data analysis, helped to conceive the computational part of the work, and helped to draft the manuscript. JMB performed the FISH experiments, DW performed the image analysis, and CB made the FISH probes. JD and JRH contributed Capture-C data on Mitoferrin1. VJB conceived and designed the experimental part of the work and helped to draft the manuscript. DM helped to conceive the computational part of the work, and helped to draft the manuscript. All authors read and approved the final manuscript. 

\section*{Acknowledgements}

We thank N. Gilbert, A. Buckle, D. Vernimmen and J. Allan for technical discussions and critical reading of the manuscript. The work was funded by EPSRC grant EP/I034661/1 and ERC Consolidator Grant 648050 THREEDCELLPHYSICS (CAB and DM), Medical Research Council MC\_UU\_12009 (JMB, CB, VJB) and Joint Research Councils MR/K01577X/1 (DW).


\bibliographystyle{bmc-mathphys} 
\bibliography{reference.bib}      



\balancecolsandclearpage

\setcounter{figure}{0}
\renewcommand{\figurename}{} 
\renewcommand{\thefigure}{\textbf{Additional file~\arabic{figure}: Figure S\arabic{figure}}}
\renewcommand{\tablename}{} 
\renewcommand{\thetable}{\textbf{Additional file~\number\numexpr\arabic{table}+12\relax: Table S\arabic{table}}}

\renewcommand{\theequation}{S\arabic{equation}}
\makeatletter
\renewcommand\tagform@[1]{\maketag@@@ {[\ignorespaces #1\unskip \@@italiccorr ]}}
\makeatother

\onecolumngrid
\begin{center}
{\huge Additional files}\\~\\
\end{center}

\begin{figure*}[h!]
\includegraphics[width=\textwidth]{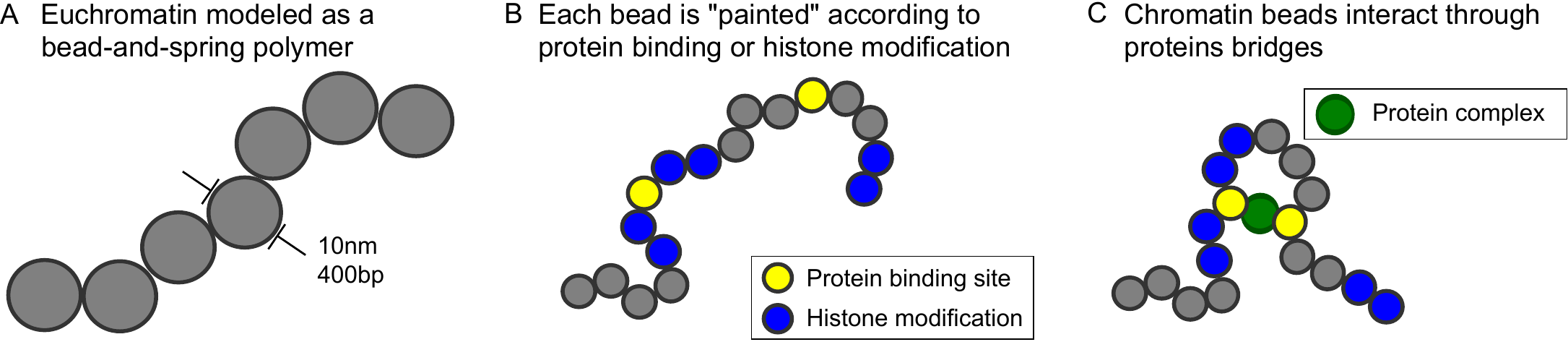}
\caption{\textbf{Chromatin is modelled as a bead-and-spring polymer.} (A) Beads represent a region of chromatin containing 400~bp of DNA, approximately two nucleosomes. This coarse graining sets the resolution of our simulations, but does not specify a particular structure for the chromatin fibre; the physical size of the bead is not specified, but fitting results to FISH measurements (see \matmeth{} and \sfref{5}) suggests a bead diameter of 16~nm. We set the persistence length (i.e. length over which the polymer behaves like a stiff rod, and a measure of the stiffness of the polymer) to 4 bead diameters (64~nm), a reasonable choice for euchromatin~(\citelangowski{}). 
Changing this parameter does not significantly affect our results (see \sfref{6}). (B) and (C) Experimental data such as ChIP-seq or DNase-seq is used to specify which beads can be bound by the protein complexes in our simulations. }\label{schematic}
\end{figure*}

\begin{figure*}
\centering
\includegraphics{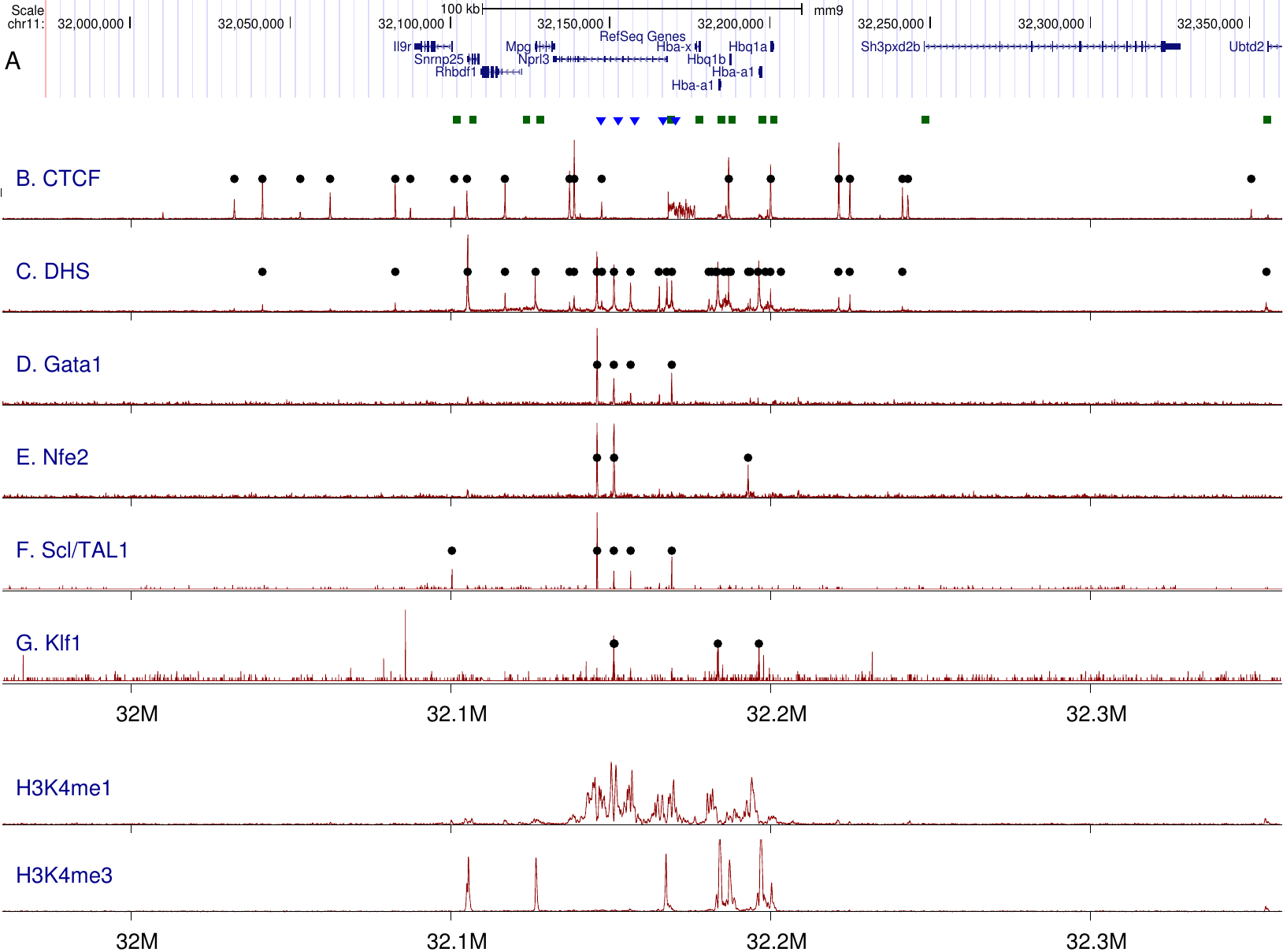}
\caption{\textbf{ChIP-seq and DNase-seq data are used as an input to the model.} (A) Genome browser view of genes in a 400~kbp region of mouse chromosome 11 surrounding the $\alpha$-globin locus which is treated in our simulations. Symbols below the browser indicate the positions of the known regulatory elements (blue triangles) and the gene promoters (green squares). 
(B) ChIP-seq data for CTCF binding across the same region from mouse erythroid (Ter119$^+$) cells. Red lines show the pile-up of reads, and black points indicate the positions of binding sites identified by peak-calling (see \supmethods{} for details). Data from Ref.~(\citeHughes{}). 
(C) Similar plot showing DNase-seq data from the same cell type, identifying the positions of DNase-1 hypersensitive sites (DHS). Data from Ref.~(\citeMarques{}). 
(D)-(G) Plots showing ChIP-seq data, again from the same cell type, for four TFs thought to be key players in globin regulation. Data from Ref.~(\citeHughes{}) 
(GATA1 and NFe2), Ref.~(\citeKassouf{}) 
(Scl/Tal1), and Ref.~(\citeTallack{}) 
(Klf1). (Note that where available, control data were used in the peak calling, meaning that peaks seen in the pile-up of reads which did not show significant enrichment above the control were not called.)  Since there are DHS located at the binding sites of each of these proteins, we reduce the complexity of our model (and the need for assumptions about the interaction between TFs) by using these as a proxy for protein binding sites. 
(H)-(I) ChIP-seq data showing relevant histone modifications: monomethylation and trimethylation of H3K4 (associated with enhances and promoters respectively). Data from Ref.~(\citeMarques{}). 
All plots are aligned according to the horizontal axis. }\label{alphaTer119inputdata}
\end{figure*}

\begin{figure*}
\centering
\includegraphics{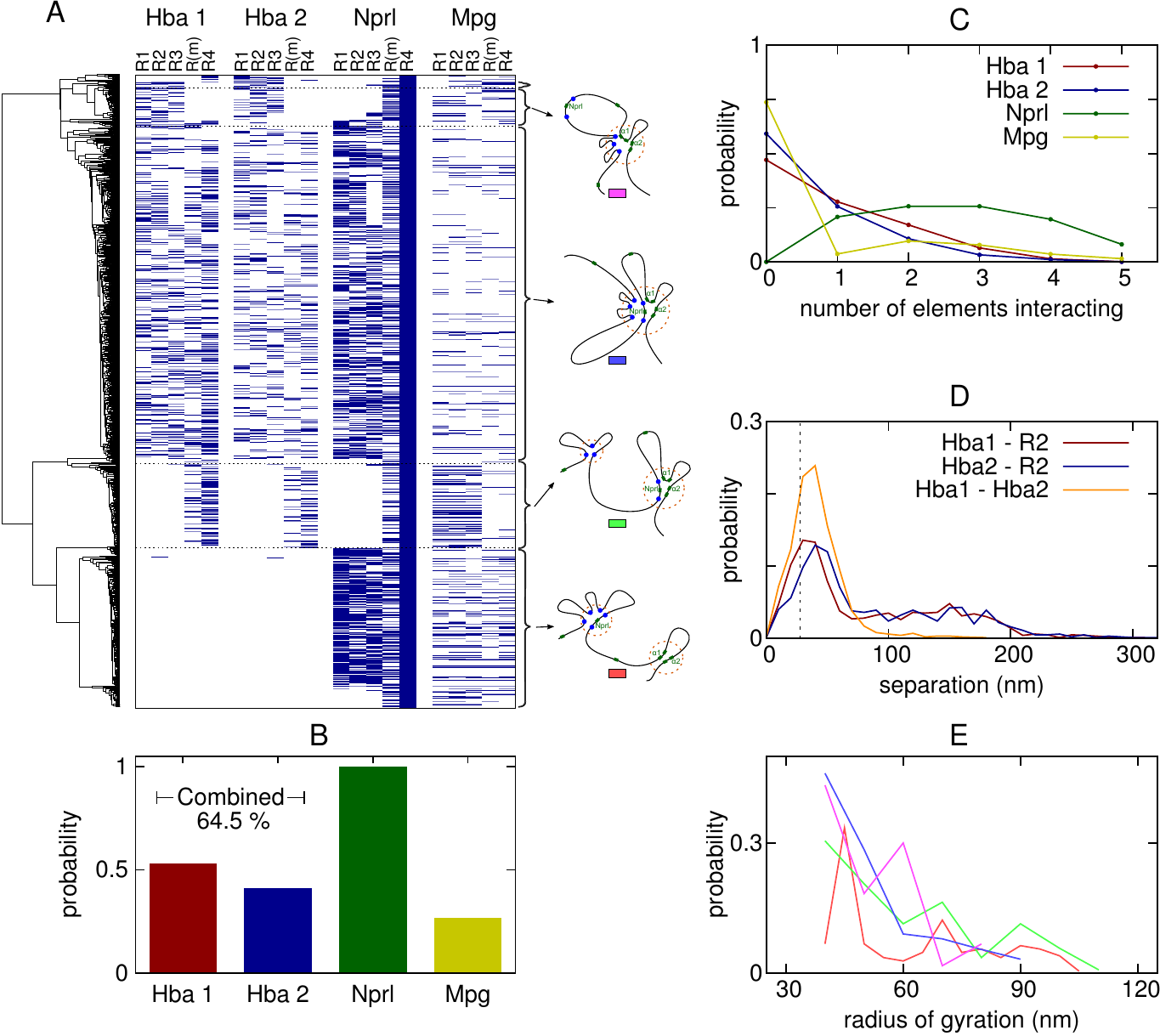}
\caption{\textbf{Interactions between promoters and specific regulatory elements can be identified in each simulated conformation.}  (A) Plot showing details of which promoters are interacting with each of the five known regulatory elements in the same set of simulations as presented in \frefs{1-3}. Each horizontal row represents a single simulated conformation, with a blue mark indicating there is an interaction with the element (an interaction is defined as any chromatin bead lying within the promoter being within 2.75 bead diameters of any chromatin bead within the regulatory element). The grouping of different types of structure according to the clustering analysis is indicated to the left. 
(B) Plot showing in what proportion of conformations each of the promoters is interacting with one or more of the regulatory elements. The proportion of conformations in which either one of the $\alpha$ globin promoters is interacting with any of the elements is also indicated; one would expect this to represent the proportion of conformation in which $\alpha$ globin is being transcribed.
(C) Histograms showing the distribution of the number of elements with which each promoter simultaneously interacts in a given conformation. 
(D) Histograms showing the distributions of the 3D separation between each of the two $\alpha$ globin gene promoters, and the R2 regulatory element. Also shown is the distribution for the separation between the two promoters. The dashed line indicates the distance below which two chromatin beads are deemed to be interacting. 
(E) Histograms showing the distribution of the radius of gyration of the locus in the conformations adopting each structure identified by the clustering analysis. The colour correspond to the different structures as shown in the schematics in panel A.
}\label{alphaTer119interactions}
\end{figure*}

\begin{figure*}
\centering
\includegraphics{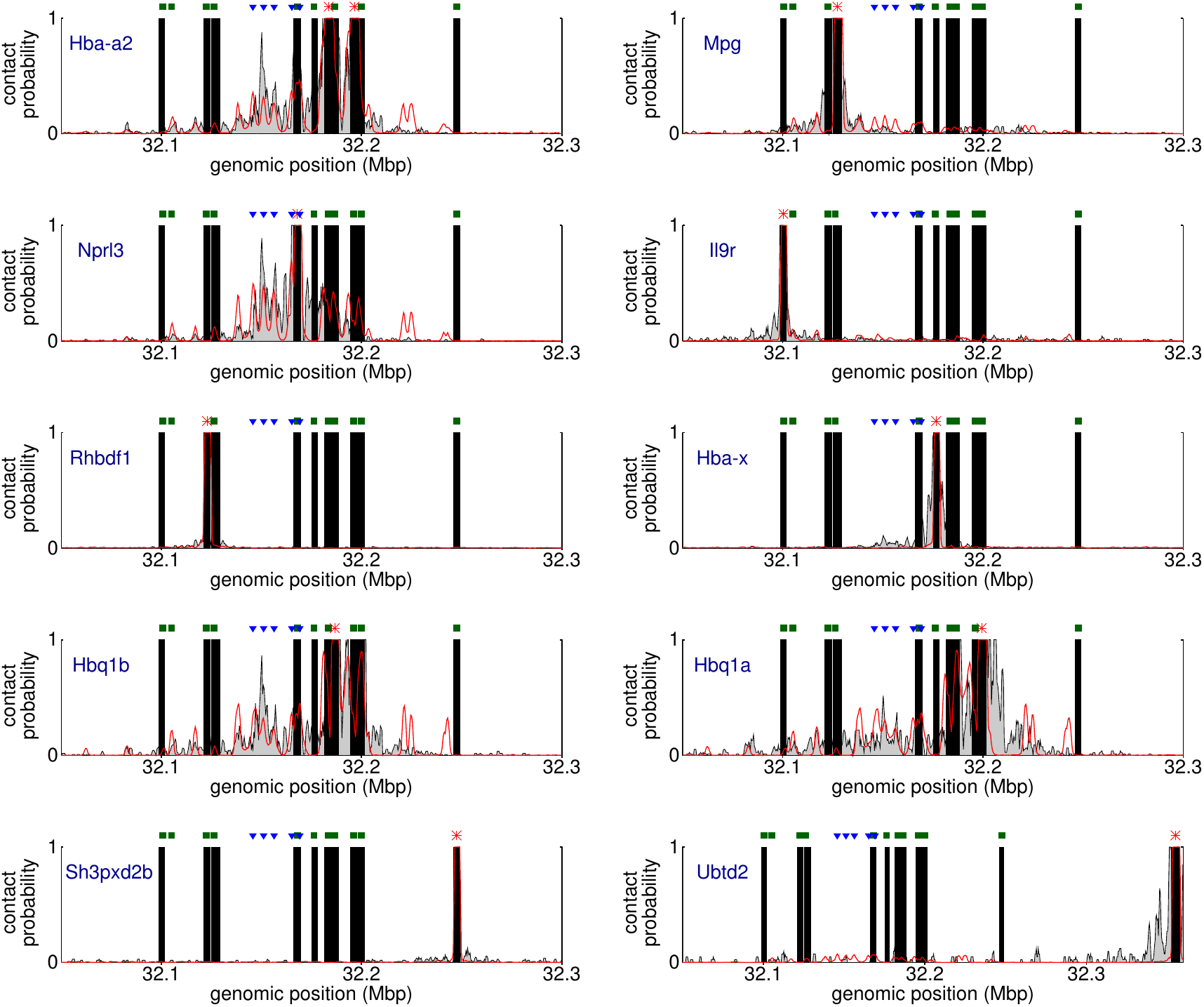}
\caption{\textbf{Capture-C data confirms many of the long range chromatin interactions within the $\boldsymbol{\alpha}$ globin locus which are predicted by simulations.} Plots showing results from the simulations described in \frefs{1-3} alongside data from Capture-C experiments. Capture-C results are from Ref.~(\citeHughes{}), 
and plots are shown for each capture probe from that data set; these are at the locations of all of the promoters within the region simulated. Red lines show simulation results, black shaded curves the experimental data, and black bars indicate regions where no experimental data is available. The experimental data are scaled as described in \supmethods{}. The positions of the known regulatory elements and other promoters are indicated at the top of each plot with blue and green symbols respectively, and the red stars indicate the position of the Capture-C probe for each plot. }\label{alphaTer119comparedata}
\end{figure*}

\begin{figure*}
\centering
\includegraphics{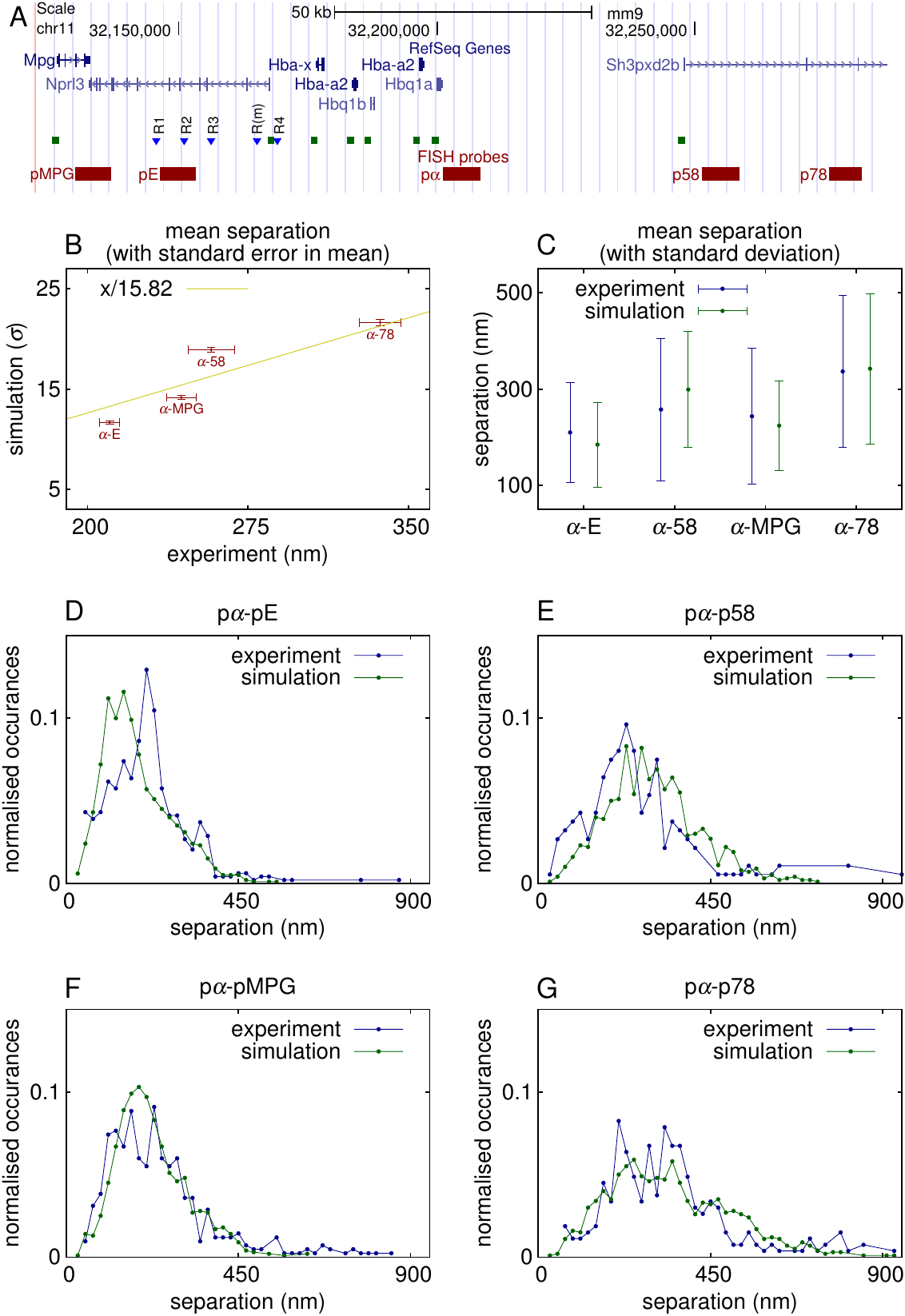} 
\caption{\textbf{Simulation results show good agreement with fluorescence in-situ hybridization measurements.} (A) Genome browser view showing the locations of FISH probes across the $\alpha$ globin locus; the positions of the known regulatory elements are indicated with blue triangles, and the promoters are indicated with green squares. All features are shown to scale.
(B) We use the mean separations of probes measured in FISH experiments to parametrise length scales in the simulations. The experiments were performed on mature erythroblasts 30~hours after differentiation, as described in \matmeth{}. Points show the experimental versus simulation mean separations with the standard error in the mean shown as error bars. The line shows a linear fit going through zero; the slope gives the conversion $\sigma=15.95$~nm, which we round to 16~nm for the rest of the plots.
(C) Plot showing the mean and standard deviation (shown as error bars) of the separation of pairs of probes.
(D-G) Plots showing the full distribution of separations across many cells (at least 187 signal pairs) or simulated conformations (1000). }\label{FISH}
\end{figure*}

\begin{figure*}
\centering
\includegraphics{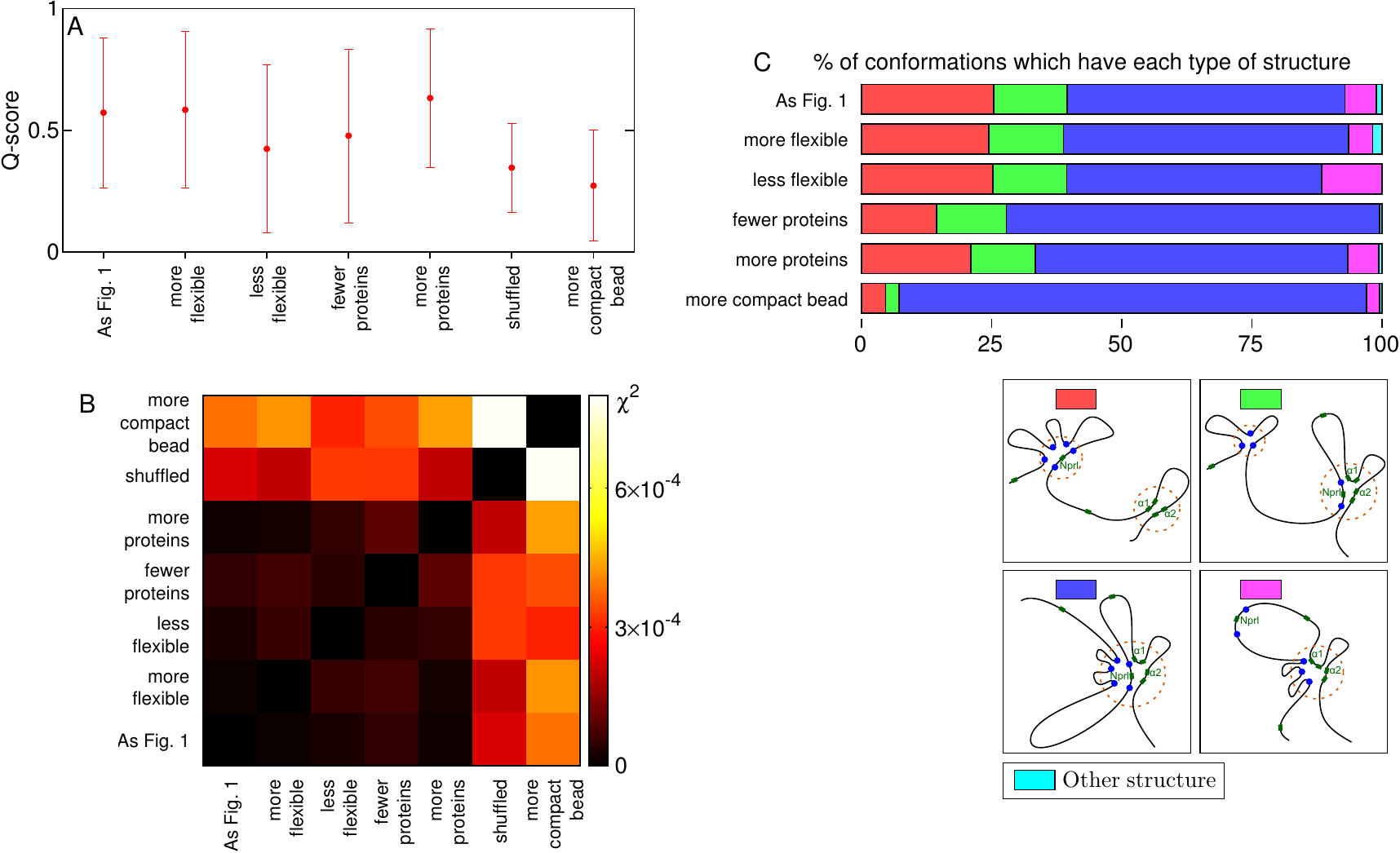} 
\caption{\textbf{Variation of model parameters does not lead to large changes in the resulting configurations.} (A) Plot showing how the $\mathcal{Q}$ score, which quantifies the agreement between the simulation and experimental chromosome interactions, varies with different model parameters (see \supmethods{}). Points show the mean over the set of targets captured in the experiment, and error bars show the error in this mean.
The flexibility of the polymer was varied (making is less flexible by increasing the persistence length from 4 bead diameters to 8 bead diameters, or making it more flexible by decreasing the persistence length to 2 bead diameters); the number of protein complexes was varied, either decreasing from 20 to 10 copies of each species, or increasing to 30 of each species. We also considered a simulation where the colouring of each bead was randomly shuffled (see \supmethods{}). Finally, we reduced the resolution of the model by increasing the amount of chromatin represented by each bead from 400~bp to 600~bp. 
(B) Plot showing how the contact maps differ between each set of experiments, measured by $\chi^2$ (see \supmethods{}).
(C) Plot showing the proportion of conformations found to be forming the different structures identified by the clustering analysis. Schematics of these structures are shown below the plot.}\label{varyparameters}
\end{figure*}

\begin{figure*}
\centering
\includegraphics{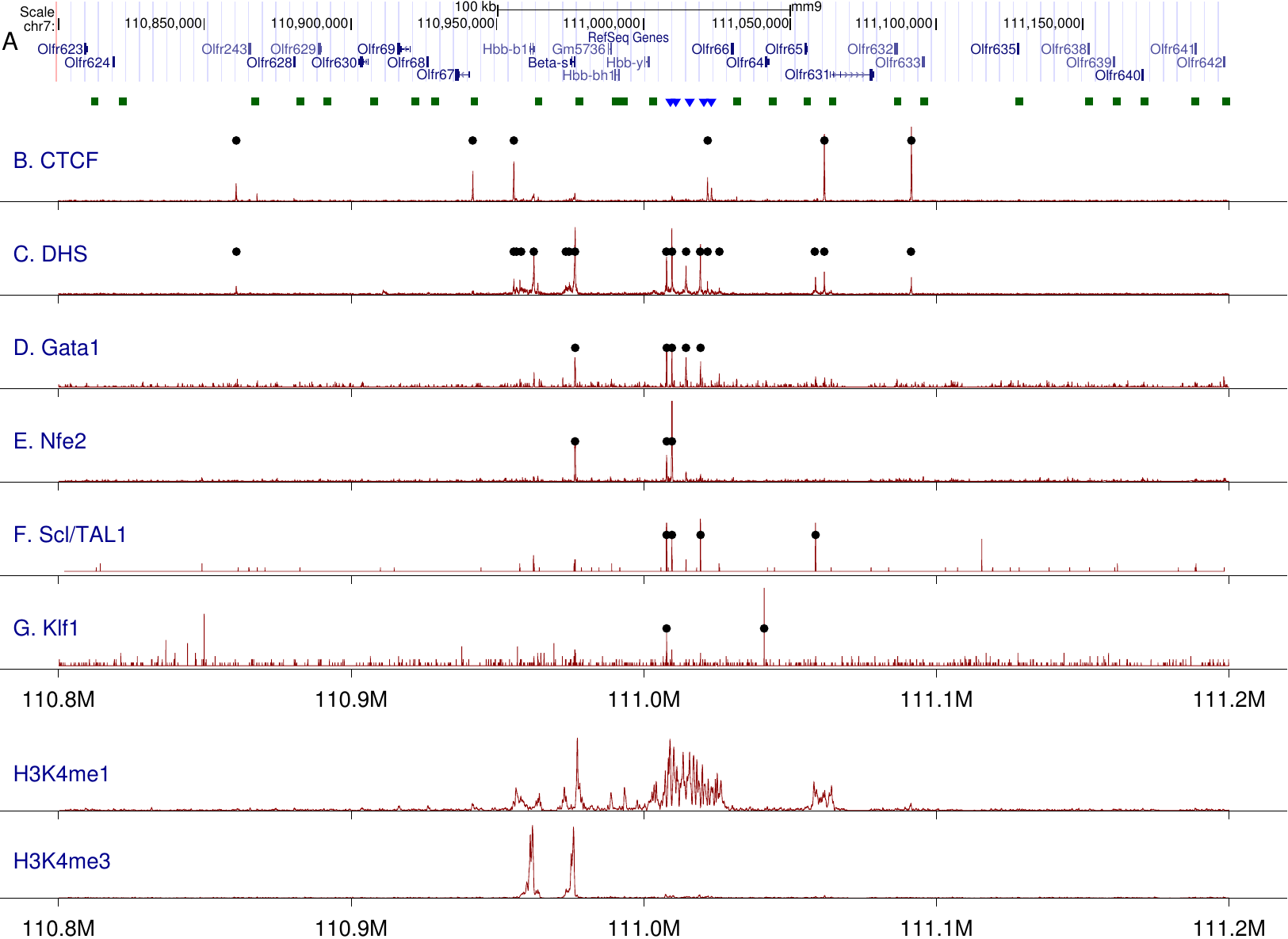} 
\caption{\textbf{ChIP-seq and DNase-seq data are used as an input to a model of the $\boldsymbol{\beta}$ globin locus.}
(A) Genome browser view of genes in a 400~kbp region of mouse chromosome 7 surrounding the $\beta$ globin locus which is treated in our simulations. Symbols below the browser indicate the positions of the known regulatory elements within the LCR (blue triangles) and the gene promoters (green squares). 
(B) ChIP-seq data for CTCF binding across the same region from mouse erythroid (Ter119$^+$) cells. Red lines show the pile-up of reads, and black points indicate the positions of binding sites identified by peak-calling (see \supmethods{}). Data from Ref.~(\citeHughes{}). 
(C) Similar plot showing DNase-seq data from the same cell type, identifying the positions of DNase-1 hypersensitive sites (DHS). Data from Ref.~(\citeMarques{}). 
(D)-(G) Plots showing ChIP-seq data, again from the same cell type, for four TFs thought to be key players in globin regulation. Data from Ref.~(\citeHughes{}) 
(GATA1 and NFe2), Ref.~(\citeKassouf{}) 
(Scl/Tal1), and Ref.~(\citeTallack{}) 
(Klf1).
(H)-(I) ChIP-seq data showing relevant histone modifications: monomethylation and trimethylation of H3K4 (associated with enhances and promoters respectively). Data from Ref.~(\citeMarques{}). 
All plots are aligned according to the horizontal axis. }\label{betaTer119inputdata}
\end{figure*}

\begin{figure*}
\centering
\includegraphics{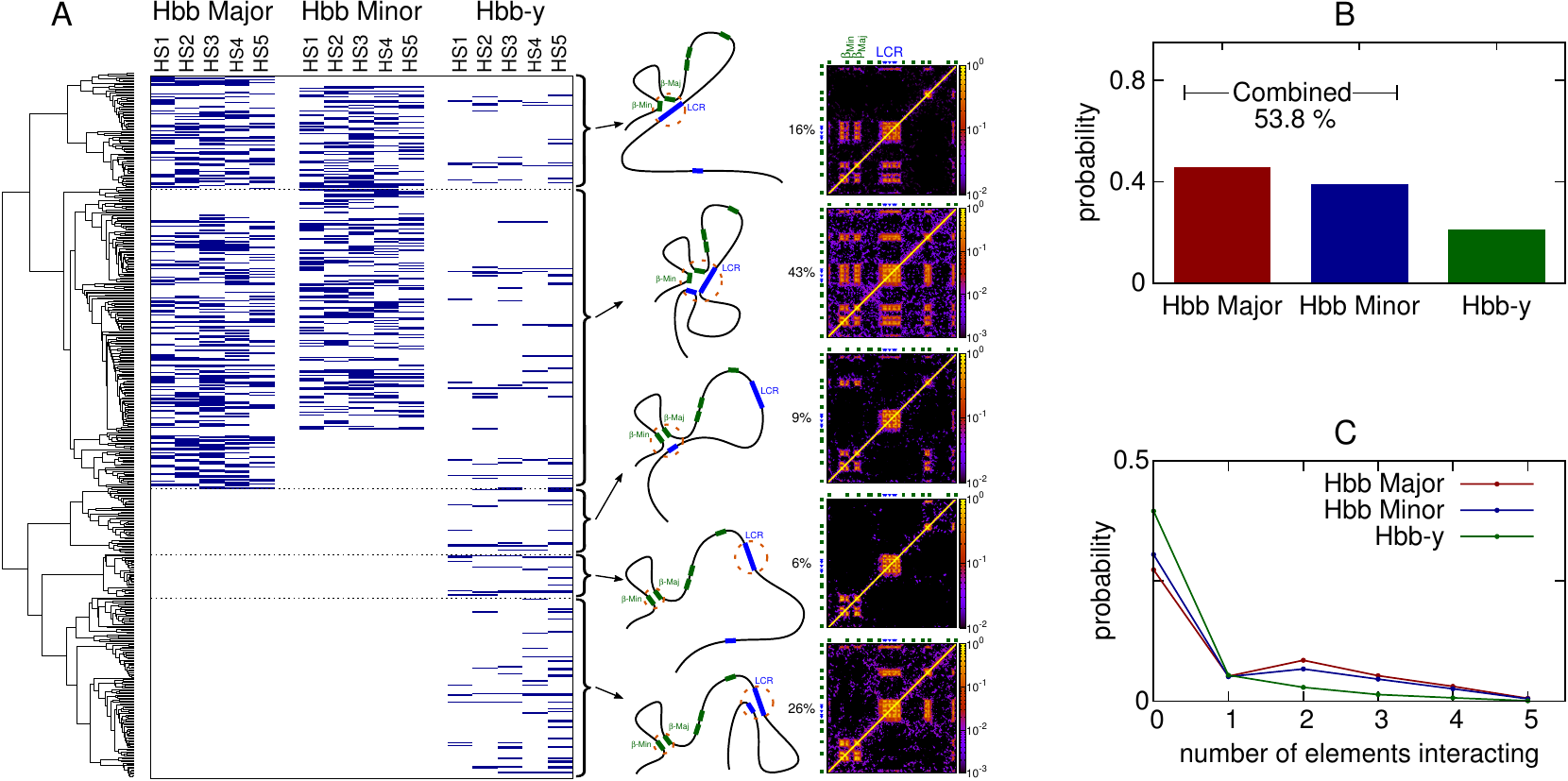} 
\caption{\textbf{Interactions between $\boldsymbol{\beta}$ globin promoters and specific regulatory elements can be identified in each simulated conformation.}  (A) Plot showing details of which promoters are interacting with the known regulatory elements within the LCR from the same set of simulations as presented in \fref{4}. Each horizontal row represents a single simulated conformation, with a blue mark indicating there is an interaction with the element (an interaction is defined as any chromatin bead lying within the promoter being within 2.75 bead diameters of any chromatin bead within the regulatory element). The grouping of different types of structure according to the clustering analysis is indicated to the left; schematics and an individual contact map for each group are shown. 
(B)  Plot showing in what proportion of conformations each of the promoters is interacting with one or more of the regulatory elements. The proportion of conformations in which either one of the $\beta$ globin promoters is interacting with any of the elements is also indicated.
(C) Histograms showing the distribution of the number of elements with which each promoter simultaneously interacts in a given conformation. 
 }\label{betaTer119elements}
\end{figure*}

\begin{figure*}
\centering
\includegraphics{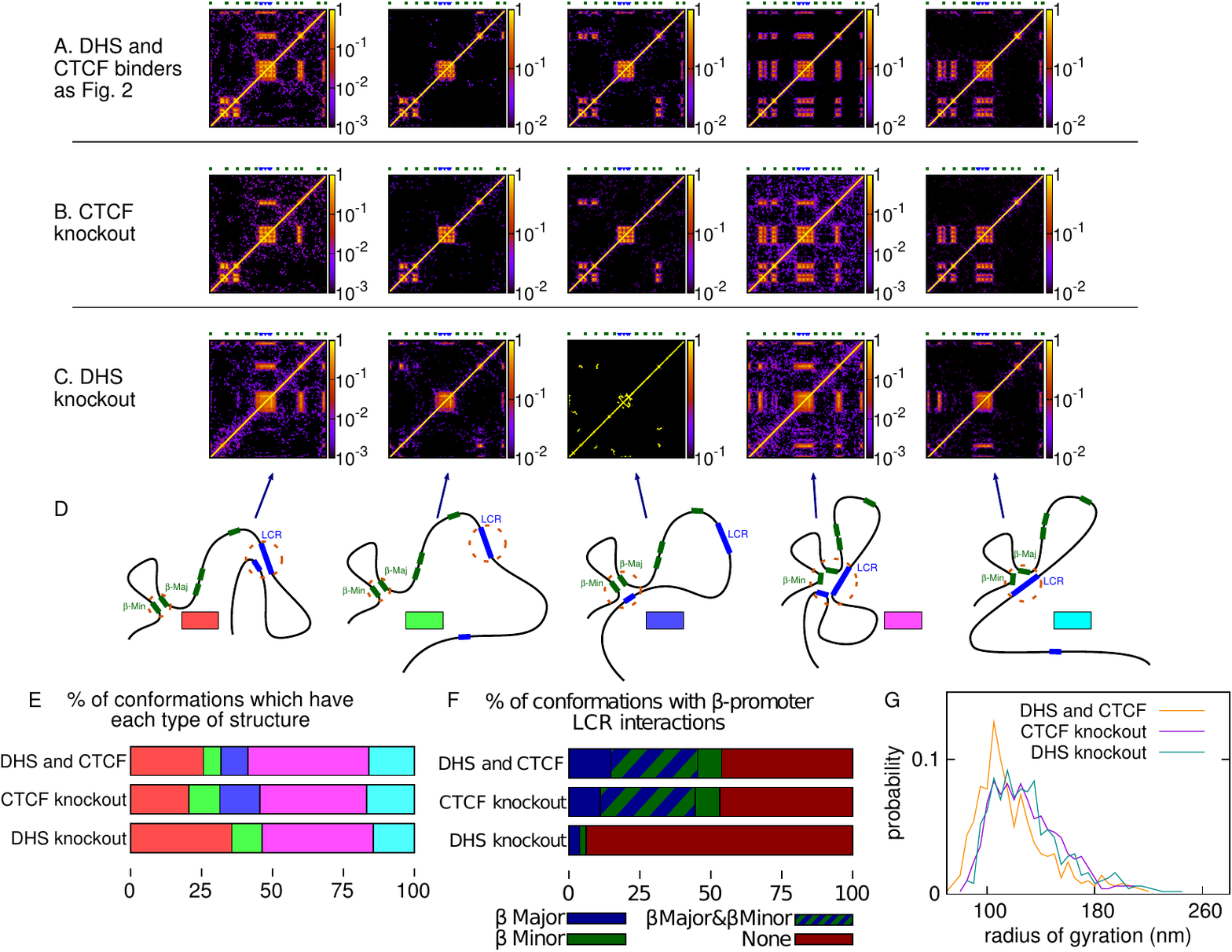}
\caption{\textbf{Simulations predict the effect of protein knock-outs on the $\boldsymbol{\beta}$ globin locus.} Plots showing the effect of a CTCF knock-out, and a ``DHS knockout'' (equivalent to knocking out all protein complexes involved in looping the $\beta$ globin locus \textit{except} CTCF). 
(A)-(C) Contact maps showing the interactions between different chromosomal locations for conformations within each group identified by clustering analysis. Maps from three sets of simulations are shown. 
(D) Schematics showing the structure of the locus within each group. 
(E) Plot showing the percentage of conformations which belong to each group identified by the clustering analysis. The colour key is given in D. 
(F) Plot showing in what percentage of conformations the two $\beta$ globin gene promoters are interacting with one or more of the regulatory elements within the LCR.
(G) Plot showing the distribution of the radius of gyration of the locus across the simulated conformations. The radius of gyration is defined as given in the caption to \fref{7}.
}\label{betaknockouts}
\end{figure*}

\begin{figure*}
\centering
\includegraphics{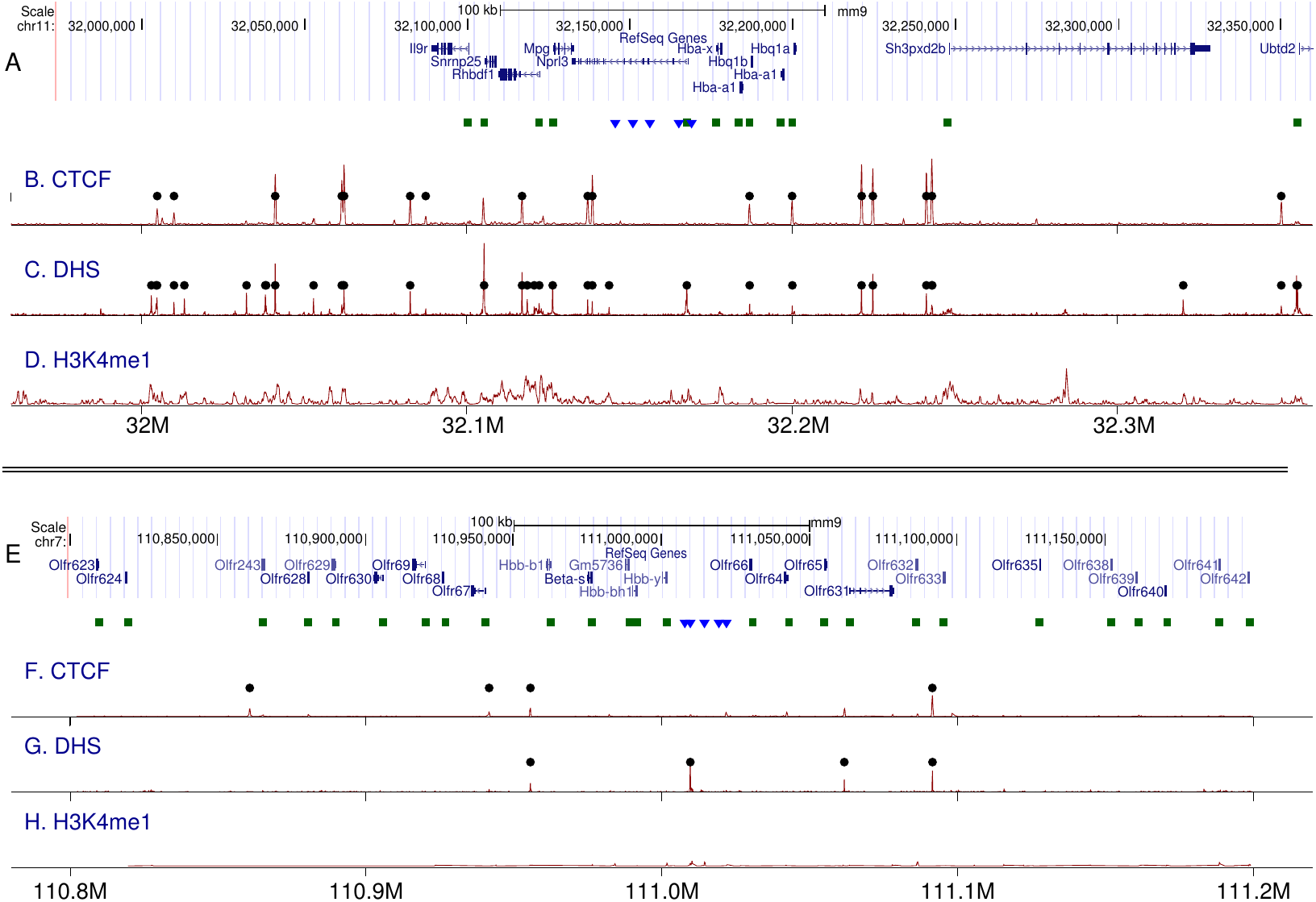} 
\caption{\textbf{ChIP-seq and DNase-seq data from mES cells can also be used as an input.} (A) Browser view of the $\alpha$ globin locus. (B-D) ChIP-seq and DNase-seq data for mouse ES cells across the same region. Red lines show the pile-up of reads, and black points indicate the positions of binding sites identified by peak-calling. Data from the ENCODE project~(\citeENCODE{}). 
(E) Browser view of the $\beta$ globin locus. (F-H) ChIP-seq and DNase-seq data for mouse ES cells across the same region. Data from the ENCODE project~(\citeENCODE{}). 
}\label{stemcellsdata}
\end{figure*}

\begin{figure*}
\centering
\includegraphics{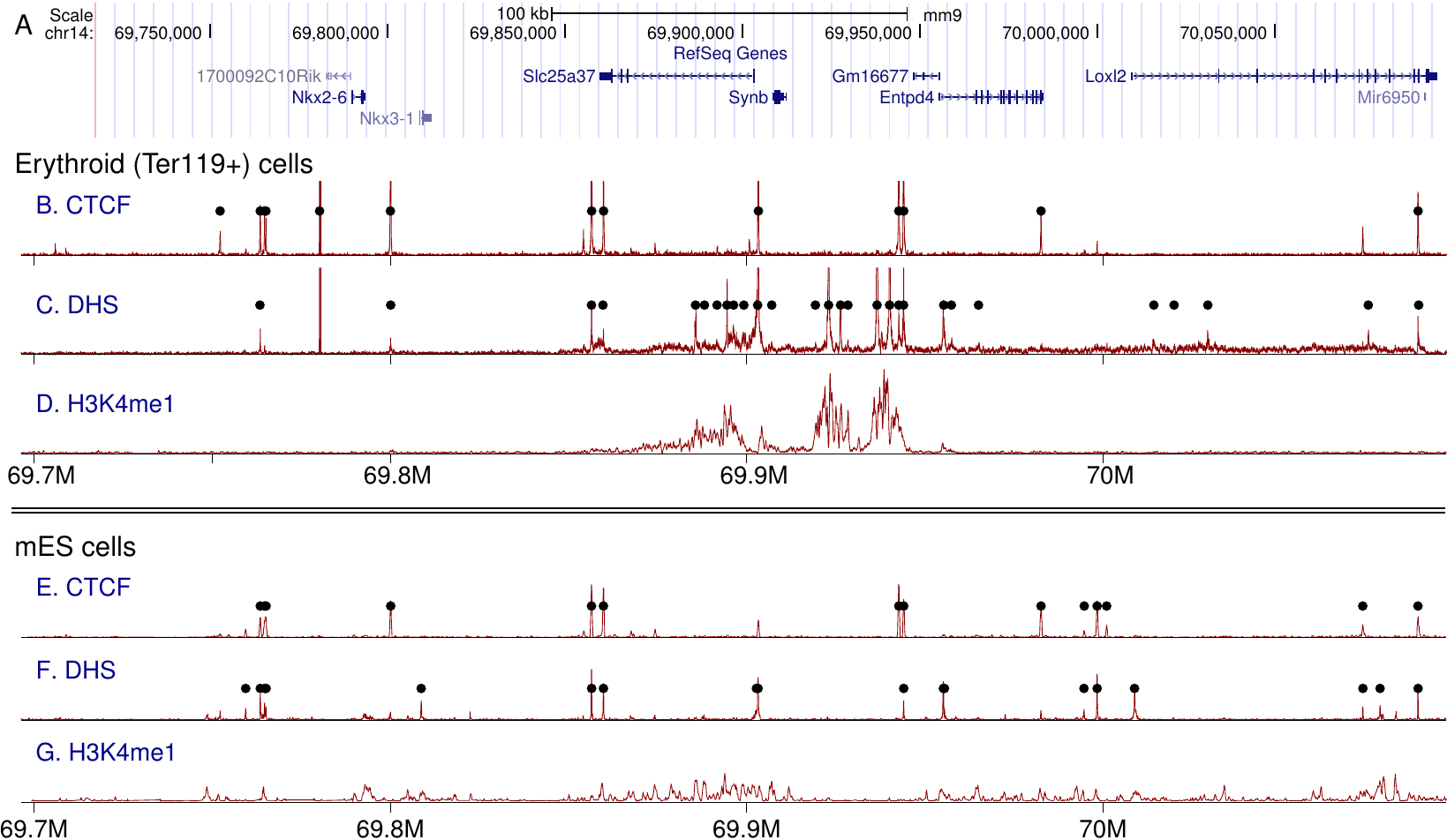} 
\caption{\textbf{Input data is available for less well studied loci.} (A) Browser view of the \textit{Slc25a37} (mitoferrin) locus. (B-D) ChIP-seq and DNase-seq data for mouse erythroid (Ter119+) cells across the same region. Red lines show the pile-up of reads, and black points indicate the positions of binding sites identified by peak-calling. CTCF data is from Ref.~(\citeHughes{}); DNase and histone modification data is from Ref.~(\citeMarques{}).
 (E-G) ChIP-seq and DNase-seq data for mouse ES cells across the same region. Data from the ENCODE project~(\citeENCODE{}). 
}\label{stemcellsdata_mito}
\end{figure*}

\begin{figure*}
\centering
\includegraphics{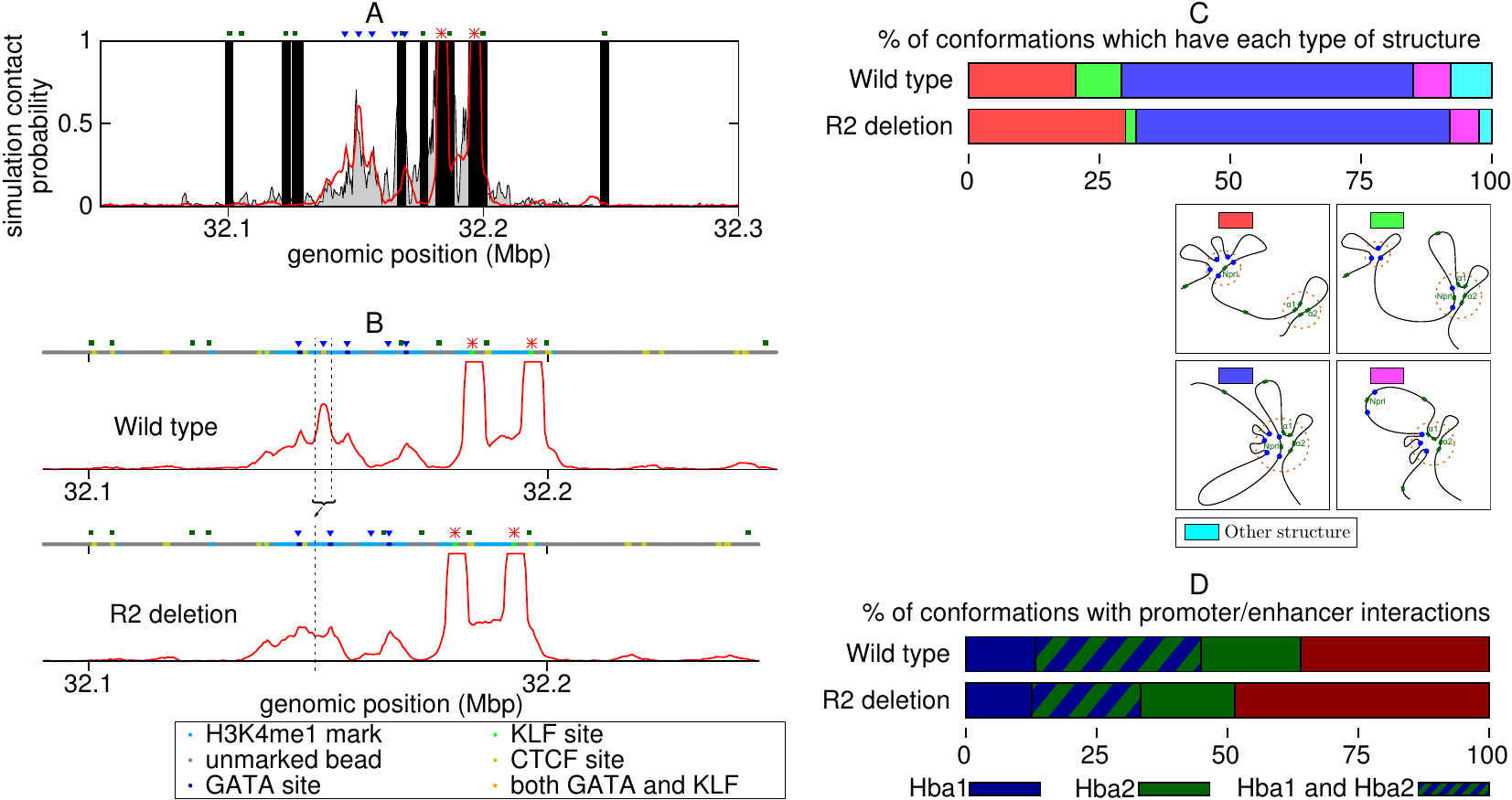} 
\caption{\textbf{A more detailed model can explain locus folding when the R2 element is deleted.}
(A) Plot comparing interactions with the $\alpha$ globin promoters in simulations with three species of bridge protein, with those from Capture-C experiments (data from Ref.~[14]). Red lines show simulation results, black shaded curves the experimental data, and black bars indicate regions where no experimental data is available. The experimental data are scaled as described in \supmethods{}. The positions of the known regulatory elements and other promoters are indicated at the top of each plot with blue and green symbols respectively, and the red stars indicate the position of the Capture-C probes. The three bridge model reproduces the differences in height of interaction peaks for the regulatory elements R1-3.
(B) Interactions for the $\alpha$ globin promoters in a wild type and a R2 knock-out simulation. Schematics indicating the binding properties of each chromatin bead are shown above the plots. The 3.6~kbp region removed in the knock-out simulation is indicated with dashed lines.
(C) Plot showing the percentage of conformations which belong to each group identified by the clustering analysis, for the wild type and R2 knock-out simulations.
(D) Plot showing in what percentage of conformations the two $\beta$ globin gene promoters are interacting with one or more of the regulatory elements within the LCR, for the wild type and R2 knock-out simulations.
}
\end{figure*}

\renewcommand{\arraystretch}{1.5}
\setlength{\tabcolsep}{10pt}


\begin{table*}
\begin{center}
\begin{tabular}{>{\centering}p{1.5cm}<{\centering}>{\raggedright}p{7cm}<{\raggedright}cc}
\textbf{Symbol} & \textbf{Description} & \multicolumn{2}{c}{\textbf{Value (simulation units)}}  \\ \hline\hline 
$\Delta t$ & time step of numerical integration of Eq.~(S1) & & $0.01~\tau_{\rm LJ}$\\ \hline
T & total run time of each simulation & & $8\times10^4~\tau_{\rm LJ}$\\ \hline
$m_i$ & mass of bead $i$ & chromatin bead & 1 \\
& & protein complex & 1 \\\hline 
$\gamma_i$ & friction coefficient & chromatin bead & 2 \\
& & protein complex & 2 \\\hline 
$d_{ii}$ & diameter of bead $i$ & chromatin bead & $\sigma$ \\
& & protein complex & $\sigma$ \\\hline 
$R_0$ & equilibrium FENE bond length for chromatin beads & & $1.6~\sigma$ \\ \hline
$K_{\rm FENE}$ & energy of the FENE bonds & & $30~k_BT$ \\ \hline
$K_{\rm BEND}$ & energy of the bending interaction for the chromatin fibre & & $4~k_BT$ \\ \hline
$r_{\rm cut}$ & cut-off for Lennard-Jones interaction between protein complex and protein binding chromatin bead & & $1.4~\sigma$ \\ \hline 
$\epsilon$ & energy for the interaction between protein complexes and binding chromatin beads & & \parbox[t][1em][c]{2.1cm}{\centering $5.39~k_BT$ \\ ($\epsilon'=10~k_BT$)} \\ \hline
$r_{\rm contact}$ & separation below which two chromatin beads are defined as being ``in contact'', i.e. we assume the beads are interacting in a way analogous to  that  which would give a signal in a 3C experiment& & \parbox[t][1em][c]{2.1cm}{\centering $2.75~\sigma$ \\ (44~nm) }
\\ \hline\hline 
\end{tabular}
\end{center}
\caption{\textbf{List of all simulation parameters.} Parameters are given in simulation units; see \supmethods{} for details of how these relate to physical units.} 
\end{table*}

\clearpage

\begin{table*}[th!]
\begin{center}
\begin{tabular}{p{0.7cm}l}
\centering \textbf{Probe} & \textbf{Oligonucleotides 5$'$-3$'$} \\ \hline\hline 
\centering pMPG & \begin{minipage}[t]{13.5cm}\raggedright 
\texttt{CTGGGGCAGACAGCCATGGTCAGTGCCCTTCCCATACTCACAGCAACCATCTGGGTGAGCgatat
        caagcttatcgataccgtcgac \\~\\
        TCAGCGAGTCGCCGGACAAGAACCTATGGGCAGTGAGTCTGCTCAGCTCAAACAGGGGCCcaccg
        cggtggagctccaatt} \end{minipage} \\ \hline 
\centering pE & \begin{minipage}[t]{13.5cm}\raggedright
\texttt{AAGCATTCAGGGCTAAGGATGTAGCTTAGTAATAGAGGCCCTGAGCTCTATGACTACCACgatat
        caagcttatcgataccgtcgac \\~\\
        GAAGATGTCTCTGAATGTTCCAAGAGTTACAGTCAGTATTTCATTTAAAAATGTACATACcaccg
        cggtggagctccaattc} \end{minipage} \\ \hline 
\centering p$\alpha$ & \begin{minipage}[t]{13.5cm}\raggedright
\texttt{CAAATTGACATGAATCAAGAATGACAACTGAGTCTTACATGGACTGTATCCAGGGTCACAgatat
        caagcttatcgatac \\~\\
        GGTATCACACACCAGGCACACATATACACATGTACGGACACATCACACACCAGGCATACATGGAC
        AGAAGcaccgcggtggagctccaat} \end{minipage} \\ \hline 
\centering p58 & \begin{minipage}[t]{13.5cm}\raggedright
\texttt{ATGGGCTTTATTCTCTCTGTCCCTCTGCAACACTGGTGTCACACAACACGAGTCTACCATCCTTA
        AAGcaccgcggtggagctccaattc \\~\\
        CGCCTGCAGCCAGTTCCCTTTTATACCTTTACCAACATGACTAGCTTCCTAAGCAGGGACATGga
        tatcaagcttatcgataccgtcgac} \end{minipage} \\ \hline
\end{tabular}
\end{center}
\caption{\textbf{Oligonucleotide sequences for FISH probes.} Sequences used to amplify the vector backbone are shown in lower case, and regions used to subclone probe sequences in upper case.}
\end{table*}

\clearpage

\raggedbottom

\onecolumngrid

\begin{center}
{\huge Additional file 15: Supporting Methods}\\~\\
\end{center}

\twocolumngrid

\normalsize

\noindent\textbf{Langevin dynamics simulations.~} Chromatin regions and protein complexes are represented by beads, and the position of the $i$th bead in the system evolves according to the Langevin equation
\begin{equation}\label{langevin}
m_i \frac{ d^2 \mathbf{r}_i }{dt^2} = -\nabla U_i - \gamma_i \frac{d\mathbf{r}_i }{dt} + \sqrt{2k_BT\gamma_i}\boldsymbol{\eta}_i(t),
\end{equation}
where $\mathbf{r}_i$ is the position of bead $i$ with mass $m_i$, $\gamma_i$ is the friction due to an implied solvent, and $\boldsymbol{\eta}_i$ is a vector representing random uncorrelated noise such that
\begin{equation}
  \langle \eta_{\alpha}(t) \rangle = 0 ~~\mbox{and}~~  \langle \eta_{\alpha}(t)\eta_{\beta}(t') \rangle = \delta_{\alpha\beta} \delta(t-t').
\end{equation}
The noise is scaled by the energy of the system, given by the Boltzmann factor $k_B$ multiplied by the temperature of the system $T$, taken to be 310~K for a cell. The potential $U_i$ is a sum of interactions between bead $i$ and all other beads, and we use phenomenological interaction potentials as described below. For simplicity we assume that all beads in the system have the same mass $m_i=m$. Equation~(\ref{langevin}) is solved in LAMMPS using a standard Velocity-Verlet algorithm.

For the chromatin fibre the $i$th bead in the chain is connected to the $i+1$th with a with a finitely extensible non-linear elastic (FENE) spring given by the potential
\begin{align}\label{FENE}
U_{\rm FENE}&(r_{i,i+1}) =& \nonumber\\ &U_{\rm WCA}(r_{i,i+1}) - 
\frac{K_{\rm FENE} R_0^2}{2} \log \left[ 1 - \left(\frac{r_{i,i+1}}{R_0}\right)^2 \right] ,
\end{align}
where $r_{i,i+1}=|\mathbf{r}_i-\mathbf{r}_{i+1}|$ is the separation of the beads, and the first term is the Weeks-Chandler-Andersen (WCA) potential
\begin{align}
\frac{U_{\rm WCA}(r_{ij})}{k_BT}  = \left\{ 
\begin{array}{ll} 
4 \left[ \left( \frac{d_{ij}}{r_{ij}}\right)^{12} - \left( \frac{d_{ij}}{r_{ij}}\right)^{6} \right] + 1, & r_{ij}<2^{1/6}d_{ij} \\
0, & \mbox{otherwise},
\end{array} \right.
\label{eq:WCA}
\end{align}
which represents a hard steric interaction which prevents adjacent beads from overlapping; here $d_{ij}$ is the mean of the diameters of beads $i$ and $j$. The diameter of the chromatin beads is a natural length scale with which to parametrize the system; we denote this $\sigma$, and use this to define all other length scales.  The second term in Eq.~(\ref{FENE}) gives the maximum extension of the bond, $R_0$; throughout we use $R_0=1.6~\sigma$, and set the bond energy $K_{\rm FENE}=30~k_BT$. The bending rigidity of the polymer is introduced via a Kratky-Porod potential for every three adjacent DNA beads
\begin{equation}\label{eq:bend}
U_{\rm BEND}(\theta)=\\K_{\rm BEND} \left[ 1 - \cos(\theta) \right],
\end{equation} 
where $\theta$ is the angle between the three beads as give by
\begin{equation}
\cos(\theta) = [\mathbf{r}_i-\mathbf{r}_{i-1}]\cdot [\mathbf{r}_{i+1}-\mathbf{r}_{i}],
\end{equation}
and $K_{\rm BEND}$ is the bending energy.  The persistence length in units of $\sigma$ is given by $l_p=K_{\rm BEND}/k_BT$. Finally, steric interactions between non-adjacent DNA beads are also given by the WCA potential [Eq.~(\ref{eq:WCA})].

Each protein complex is represented by a single bead and the WCA potential is used to give a steric interaction between these. Chromatin beads are labelled as binding or not-binding for each protein species according to the input data (see section on ChIP-seq and DNase-seq data analysis below). For the interaction between proteins and the chromatin beads labelled as binding, we use a shifted, truncated  Lennard-Jones potential
\begin{equation}
U_{\rm LJcut}(r_{ij})= \left\{ \begin{array}{cl} 
U_{\rm LJ0}(r_{ij}) - U_{\rm LJ0}(r_{\rm cut}) & r_{ij}<r_{\rm cut}, \\ 
0 & \mbox{otherwise}, \end{array}\right. 
\label{eq:LJ}
\end{equation}
with
\[
U_{\rm LJ0}(r)=4\epsilon' \left[ \left( \frac{d_{ij}}{r}\right)^{12} -  \left( \frac{d_{ij}}{r}\right)^{6} \right],
\]
where $r_{\rm cut}$ is a cut off distance, and $r_{ij}$ and $d_{ij}$ are the separation and mean diameter of the two beads respectively. This leads to an attraction between a protein and a chromatin bead if their centres are within a distance $r_{\rm cut}$. Here $\epsilon'$ is an energy scale, but due to the second term in Eq.~(\ref{eq:LJ}) this is not the same as the minimum of the potential, which for clarity we denote $\epsilon$ (and we refer this to as the interaction energy). For simplicity we set the diameter of the protein complexes equal to that of the chromatin beads, $d_{ij}=\sigma$, and set $r_{\rm cut}=1.4~\sigma$.

The polymer is initialized as a random walk, and the dynamics are first evolved in the absence of protein interactions in order to generate an equilibrium coil conformation. Interactions with the protein complexes are then switched on, and the dynamics are evolved until a new equilibrium conformation is obtained. The length scale $\sigma$, mass $m$ and energy scale $k_BT$ give rise to a natural simulation time unit $\tau_{\rm LJ}=\sqrt{\sigma^2 m /k_BT}$, and Eq.~(\ref{langevin}) is integrated with a constant time step $\Delta t=0.01\tau_{\rm LJ}$, for a total of at least $8\times10^6$ time steps. Each simulation is repeated at least 500 times using a different initial conformation and random noise, resulting in an ensemble of conformations. Two chromatin beads are said to be interacting if their separation is less than 2.75 bead diameters; counting the proportion of conformations in which a given pair of beads is interacting gives an approximation of the probability that those beads interact.

So far the system has been described in units $\sigma$, $m$, and $k_BT$. In order to map these simulation units to real ones we must recognise that there are two further important time scales in the system, namely the inertial time $\tau_{\rm in}=m/\gamma_i$ (from Eq.~(\ref{langevin}) this is the time over which a bead loses information about its velocity), and the Brownian time $\tau_{\rm B}=\sigma^2/D_i$ (the time it takes for a bead to diffuse across its own diameter $\sigma$). Here $D_i$ is the diffusion constant for bead $i$, given through the Einstein relation by $D_i=k_BT / \gamma_i$; if we make the approximation that a chromatin bead will diffuse like a sphere we can then use Stokes' Law, where $\gamma_i=3\pi\nu d_i$, with $\nu$ the viscosity of the fluid, and $d_i$ the diameter of bead $i$. Taking realistic values for the length, mass and viscosity one finds that $\tau_{\rm in} \ll \tau_{\rm LJ} \ll \tau_{\rm B}$, with the times separated by several orders of magnitude. For numerical stability we must choose the time step $\Delta t$ smaller than all of these times, and we wish to study phenomena which will occur on times of the order $\tau_{\rm B}$; this means that using real values for all parameters would lead to infeasibly long simulation run times. Instead we make an approximation by setting $m=k_BT=\sigma=1$, and $\gamma_i=2$, and map to real time scales through the Brownian time $\tau_{\rm B}$; although this means that beads in our simulation have more inertia than in reality, this does not effect our results, which are taken once the polymer has reached an equilibrium conformation.
Taking the diameter of the chromatin beads to be $15.8$~nm, and assuming a viscosity of $10$~cP for the nucleoplasm gives $\tau_{\rm B}\approx 87~\mu$s, meaning that a simulation time unit is $\approx43.5~\mu$s. Each simulation run therefore represents approximately 7~s of real time.

\bigskip

\noindent\textbf{ChIP-seq and DNase-seq data analysis.~}
As an input to the model we use ChIP-seq and DNase-seq data (previously published in Refs.~(\citeHughes{},\citeKassouf{},\citeMarques{}-\citeENCODE{}) 
as indicated in the captions for Additional files 2, 7 and 10: Figures S2, S7 and S10) 
to identify protein binding sites in the chromosome region of interest. For protein binding, ChIP-seq reads are aligned to the mouse reference genome build mm9 using the Bowtie2 software~\cite{bowtie2}; duplicate reads are removed, and pile-ups are generated using the BedTools package~\cite{bedtools}. Binding sites are identified using the macs2 peak calling software~\cite{macs2} using a control data set where available; peaks which have a normalised $p$-value $<0.001$, and which have a fold-change higher than a threshold are retained. DNase-seq reads are similarly aligned to the mm9 genome using Bowtie2, but peaks are identified using the PeaKDEck software (which uses a peak finding algorithm calibrated specifically for DNase-seq data~\cite{peakdeck}). As detailed in the main text, we simplify our model by assuming that DNase hypersensitive sites indicate the positions of transcription factor binding sites. For histone modifications, we also align reads using Bowtie2; since these modifications can be found across wide regions, rather than identifying peaks we instead find regions where the pile-up of reads exceeds a threshold.

In order to incorporate the data into the simulations, the locus of interest is divided into regions corresponding to each bead in our model chromatin fibre. Beads are then labelled according to any peak or histone modification which overlaps with the region; for simplicity we only label beads a binding or not (or as having a histone modification or not), and do not incorporate peak intensities into the model.

\bigskip

\noindent\textbf{Cluster Analysis.~}
In order to assess the similarity between the conformations generated in each set of simulations we perform a cluster analysis. First we calculate the generalised ``distance'' between all pairs of conformations; then a dendrogram is generated using the standard hierarchical clustering algorithm in the MATLAB software~\cite{matlab}, with an average linkage criterion.

A standard way to measure the distance between two polymer conformations is to consider the mean squared difference between separations of pairs of beads in each; however since our polymer consists of regions which bind proteins and unstructured regions, this does not perform well (the unstructured regions dominate in the mean, and no clear clusters are found). Instead we use a distance $\Gamma(C,C')$ between conformations $C$ and $C'$ which ignores the unstructured regions, defined as
\begin{equation}
\Gamma(C,C') = \frac{1}{(n(n-1))/2} \sum_{i\neq j} [ 1-\delta_{s_{ij}^C,s_{ij}^{C'}} ] ( r_{ij}^C - r_{ij}^{C'} )^2,
\end{equation} 
where $r_{ij}^C$ is the separation of beads $i$ and $j$ in conformation $C$. The Kronecker $\delta$-function is defined such that $\delta_{a,b}=1$ if $a=b$ and 0 otherwise, with $s_{ij}^C=1$ if beads $i$ and $j$ are interacting in conformation $C$ and 0 otherwise (an interaction is defined as having separation less than 2.75 bead diameters). Thus the only contributions to the mean are from beads which are interacting in one conformation but not in the other; further limiting the analysis to consider only the chromatin beads within the most structured region of the locus (indicated by green bars in \freftwo{1C}{4C}) results in a series of well defined clusters (\freftwo{2}{4D}). 

\bigskip

\noindent\textbf{Capture-C data.~}
In order to test the predictions of the model we compared simulation results with Capture-C data; for the $\alpha$ and $\beta$ globin loci data were from Ref.~\citeHughes{}, whereas data for the mitoferrin locus in \fref{6} were from new experiments performed according to the method in that reference. In these experiments
a set of oligonucloetide capture ``targets'' is designed, a 3C library is obtained using a frequently cutting restriction enzyme (Dpn II, cutting at GATA), and SureSelect oligonucleotide capture is followed by Hi-seq paired-end sequencing. The resulting reads then undergo \textit{in silico} DpnII digestion (producing a set of fragments for each read), and the fragments are aligned to the mouse mm9 reference genome as single-end reads using the Bowtie software~\cite{bowtie}. Identical sets of read fragments are assumed to be PCR artefacts, and are removed~(\citeHughes{}); read sets which contain a targeted restriction fragment and a reporter fragment are retained. Data are then smoothed by counting interactions within 800~bp windows centred on genomic positions separated by 400~bp steps, giving an interaction profile for each target (black lines with grey shading in \freftwo{3A,B,~4E,F}{5C,F}, and Additional Files 4 and 12: \sfreftwo{4}{12}). Since the efficiency of capture of each target is unknown, the obtained profiles show \textit{relative} interaction strength, and profiles from different targets cannot be compared quantitatively. Reads showing interactions between targeted regions could have been captured from either target, so these reads are not quantitative and must be removed; these regions are indicated by black blocks in \freftwo{3A,B,~4E,F}{5C,F}, and Additional Files 4 and 12: \sfreftwo{4}{12}.

To compare Capture-C data with our simulated interaction profiles we first identify the simulation beads that correspond to each of the targeted regions. From the ensemble of simulated conformations we find the probability that that any chromatin bead within the target region is interacting (separation less than 2.75 bead diameters) with each other bead (the probability is approximated by $n/N$
when there is an interaction in $n$ conformations in a set of $N$). Since the Capture-C experiment only gives relative interaction profiles, to plot the data on the same axis as simulations we must scale it by a factor $\gamma$ which we find via a least squares fit. After removing interactions between targets from both the simulation and experimental data sets, we use cubic spline interpolation to obtain points at the same genomic locations for each data set; in a plot with simulation and experimental values on the axes, $\gamma$ is this slope of a linear fit which goes through zero.

\bigskip

\noindent\textbf{Quantitative comparison with experimental data - the $\mathcal{Q}$ score.~} In order to quantitatively compare our simulations with data from Capture-C experiments we define a score, denoted $\mathcal{Q}$ which takes a value between 0 and 1 depending on the overlap between chromatin interaction peaks which are predicted by simulations, and those observed in experiments ($\mathcal{Q}=1$ denoting perfect overlap). For a data set for a given capture target we first normalise by dividing by the number of interactions in the vicinity of the target; we then scale all of the experimental data so that it best fits the simulation. A sliding averaging window is used to smooth both the simulation and experimental data, before applying a peak finding algorithm to identify interactions (the ``findpeaks'' function in the MATLAB software~\cite{matlab}). We use the peak positions and widths (but not heights) to test whether peaks in each data set overlap, and calculate a value
\begin{equation}
q_i = \frac{ n_{\rm se} + n_{\rm es}  }{ n_{\rm s} + n_{\rm e} },
\end{equation}
where $n_{\rm s}$ and $n_{\rm e}$ are the number of peaks found in the simulation and experimental data respectively, $n_{\rm se}$ is the number of peaks in the simulation data which overlap with one or more peaks in the experimental data, and $n_{\rm es}$ is the number of peaks in the experimental data which overlap with one or more peaks in the simulation data. It is possible for $n_{\rm se}$ and $n_{\rm es}$ to differ if, for example, two adjacent peaks in the simulation overlap a single broader peak in the experiment. Since from a single simulation and experiment we compare data from each capture target separately, we take an average to find an overall score $\mathcal{Q}=\sum_i q_i$, where $q_i$ is the score for the $i$th capture target. Note that since the experimental data is always scaled so as to best fit the simulation (necessary since the Capture-C signal is in units of numbers of reads, and we do not know the proportionality constant which relates this to the probability of two regions interacting), simulations always score reasonably well, and defining a measure of their quality is very difficult. To set the scale, we compare with a simulation where the bead colourings are shuffled randomly.

In \sfref{6} we compare $\mathcal{Q}$ scores for a number of different simulation models. To generate the ``shuffled'' chromatin fibre, the bead colourings are shuffled subject to two constraints: first, in order to preserve e.g. the pattern of histone methylation around the DHS or CTCF sites, we keep groups of 10 adjacent beads (4~kbp) together, and second, so that there are some interactions to compare we preserve the bead colouring at the targets used in the experiment (if a protein binding site were shuffled away from a target, then there would be very little long range interaction with that region).

Quantifying the difference between two different sets of simulations is more straightforward, since no scaling is required. We define 
\vspace{-0.5cm}
\begin{equation}
\chi^2(A,B) = \frac{1}{(n(n-7))} \sum_{|i-j|>6} [ P_A(i,j)-P_B(i,j) ]^2,
\end{equation}
where $P_A(i,j)$ is the probability that chromatin beads $i$ and $j$ are in contact in set of simulations $A$ (i.e. the values shown in contact maps), and the sum runs over all pairs of beads which have a linear separation greater than 6 (this means the diagonal in contact maps are not included in the comparison). $\chi^2$ gives the difference between two contact maps, i.e. the larger its value the more different the two sets of experiments.

\let\oldthebibliography=\thebibliography
\let\oldendthebibliography=\endthebibliography
\renewenvironment{thebibliography}[1]{%
  \oldthebibliography{#1}%
  \setcounter{NAT@ctr}{58}%
}{\oldendthebibliography}


\end{document}